\documentclass[journal, final]{new-aiaa}
\usepackage[utf8]{inputenc}
\usepackage{textcomp}

\usepackage{amsmath}
\usepackage[version=4]{mhchem}
\usepackage{siunitx}
\usepackage{longtable,tabularx}
\usepackage{breqn}
\usepackage{comment}
\usepackage{mathtools}
\usepackage{graphics}
\usepackage{algorithmic}
\usepackage{tikz}
\usetikzlibrary{automata, positioning}

\usepackage{todonotes}
\newcommand{\question}[2][]{\todo[color=green!20, #1]{#2}}

\DeclareMathOperator*{\argmax}{arg\,max}
\DeclareMathOperator*{\argmin}{arg\,min}

\newtheorem{proposition}{Proposition}[section]

\title{Safety-Aware Hybrid Control of Airborne Wind Energy Systems}

\author{Nikolaus Vertovec \footnote{Doctoral Student, Department of Engineering Science, University of Oxford; nikolaus.vertovec@eng.ox.ac.uk.}}
\affil{Department of Engineering Science, University of Oxford
Parks Road, OX1 3PJ}

\author{Sina Ober-Bl\"obaum \footnote{Professor, Department of Mathematics, Paderborn University,
33100; sinaober@math.uni-paderborn.de}}
\affil{Department of Mathematics, Paderborn University,
33100}

\author{Kostas Margellos \footnote{Associate Professor, Department of Engineering Science, University of Oxford; kostas.margellos@eng.ox.ac.uk}}
\affil{Department of Engineering Science, University of Oxford
Parks Road, OX1 3PJ}

\begin{document}

\maketitle

\begin{abstract}
A fundamental concern in progressing Airborne Wind Energy (AWE) operations towards commercial success, is guaranteeing that safety requirements placed on the systems are met. Due to the high dimensional complexity of AWE systems, however, formal mathematical robustness guarantees become difficult to compute. We draw on research from Hamilton-Jacobi (HJ) reachability analysis to compute the optimal control policy for tracking a flight path, while enforcing safety constraints on the system. In addition, the zero-sublevel set of the computed value function inherent in HJ reachability analysis indicates the backward reachable set, the set of states from which it is possible to safely drive the system into a target set within a given time without entering undesirable states. Furthermore, we derive a switching law, such that the safety controller can be used in conjunction with arbitrary least restrictive controllers to provide a safe hybrid control law. In such a setup, the safety controller is only activated when the system approaches the boundary of its maneuverability envelope. Such a hybrid control law is a notable improvement over existing robust control approaches that deteriorate performance by assuming the worst-case environmental and system behavior at all times. We illustrate our results via extensive simulation-based studies.
\end{abstract}

\section{Introduction}
With the looming threat of the climate crisis becoming ever more detrimental, the need for low-cost and reliable sources of energy becomes increasingly pressing. Wind remains the primary non-hydro renewable and to reach net-zero energy targets of 8008 TWh in 2030, will need to grow annually by $18\%$ between 2021 and 2030 \cite{IEA_2021}. Conventional wind turbines have high material costs and often struggle with unreliable wind patterns present at low altitudes. Airborne Wind Energy (AWE) systems solve many of the problems of conventional wind turbines by harnessing wind energy at high altitudes, where stronger and more reliable wind currents can be found. Thus, over the past decade, we have seen a rapid increase in research and development into AWE from both academia and the private sector \cite{Schmehl2018, AWE2018}.

In this paper, we only consider Ground-Gen systems, whereby energy is obtained by continuously performing two phases of flight, a \textit{traction} phase, and a \textit{retraction} phase. During the traction phase, a tethered kite or fixed-wing aircraft is flown in crosswind conditions at altitudes of up to 1000m above ground. The traction force acting on the aircraft's tether is converted into electricity using a generator and winch located at the base of the tether \cite{Fagiano2014, Cherubini2015}. For a discussion of alternative AWE systems, we refer to \cite{Cherubini2015, Vermillion2021}. Since the tether is gradually reeled out during the traction phase, the system will eventually need to be reset. For this, the aircraft is flown back to its starting altitude upwind in a phase commonly referred to as the retraction phase. During the retraction phase, a small amount of the previously generated energy needs to be spent on reeling in the tether. Nevertheless, AWE systems have been shown to be able to create a net energy output of over 20 kW \cite{Cherubini2015} with proposed larger systems producing energy in the range of MWs \cite{Eijkelhof2020}.

\begin{figure}
 	\centering
	\includegraphics[width=.6\textwidth]{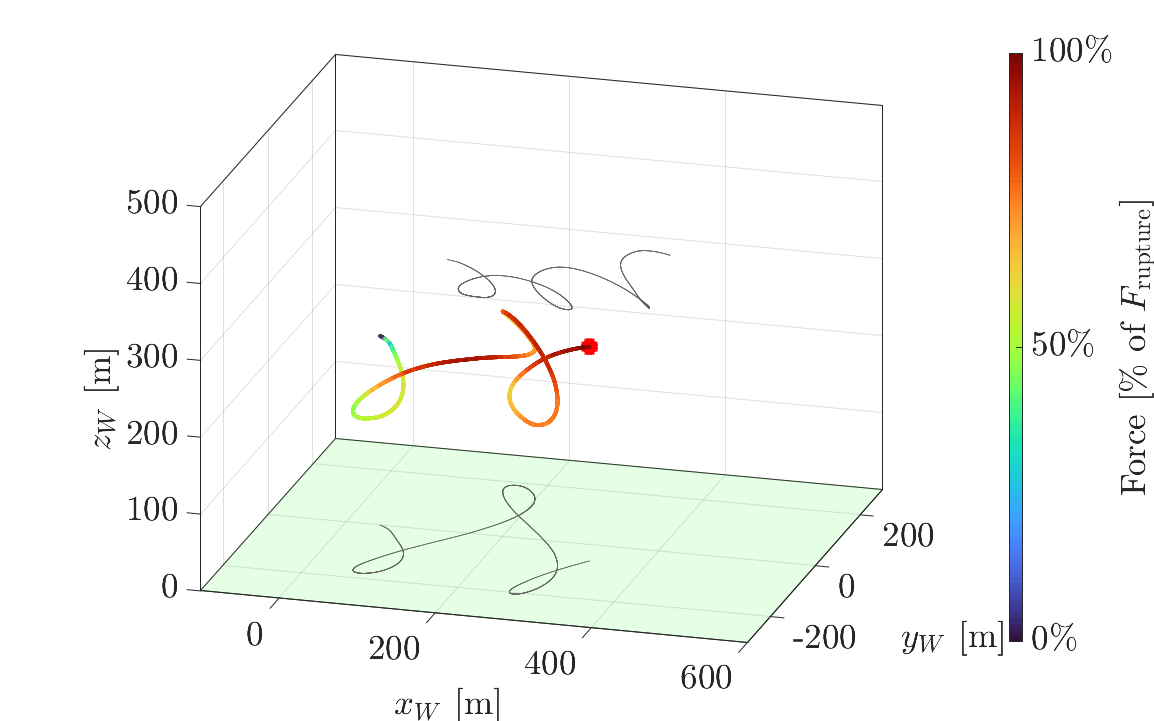}
	\caption{Tether rupture during traction phase of flight. The red dot marks the point during flight at which a tether rupture occurs due to the maximum tether force being exceeded.}
	\label{fig:tether_rupture}
\end{figure}

For AWE systems to become widely adopted, it is necessary to provide formal guarantees on the state operation and control of such systems. The aim of this paper is to ensure robustness in the sense of safety-aware control of an AWE system with minimal impact on power generation. To this end, we propose a hybrid control law, whereby a safety controller is only activated when a critical system failure is predicted to occur. We consider an AWE setup as proposed in \cite{Rapp2021}, whereby a fixed-wing aircraft is flown in a figure of eight flight pattern, controlled using nonlinear dynamic inversion (NDI), a control strategy common in aviation applications. The tether configuration is changed from \cite{Rapp2021}, such that a tether rupture is deemed to occur when a maximum tether force of 1.87 kN is exceeded. As can be seen in Fig. \ref{fig:tether_rupture}, the tether ruptures before the aircraft can complete one full cycle of flight. Using Hamilton Jacobi (HJ) reachability analysis, we will synthesize a safety controller as well as switching conditions, such that in a hybrid control setup, minimal penalties on power generation are incurred, while ensuring a safe flight without tether rupture.

HJ reachability analysis has become a well-adopted formal verification method, whereby the optimal controller is derived through computation of the backward reachable set (BRS). Its advantages include compatibility with nonlinear system dynamics, formal treatment of nonlinear bounds, and recently also with multi-objective optimization problems \cite{Bansal2018, Vertovec2021}. However, since HJ reachability analysis requires solving a quasi-variational inequality over a gridded state space, it is subject to the curse of dimensionality. As such, its applications have been limited to low-dimensional systems \cite{Chen2018}.

To this end, we will derive a novel low-dimensional formulation of the system dynamics of AWE systems and formulate the safety critical control problem as a differential game of two players. Furthermore, for numerical reasons that will become evident later on, we are required to grid the state space, thus necessitating a coordinate transformation that ensures only relevant states are considered for controller synthesis, minimizing numerical overhead. Thus the key contributions of this paper are
\begin{enumerate}
    \item the development of a simplified low dimensional AWE model that is accurate enough for safety-critical control purposes (Section \ref{sec:modelforsafety}),
    \item the synthesis and deployment of the safety-critical controller based on the developed abstractions and HJ reachability analysis, a non-trivial application for which we have made the code and simulation environment available (Section \ref{sec:HJ}),
    \item the computation of the optimal control/disturbance actions that optimize the Hamiltonian of the underlying optimal control problem, which is case-dependent and non-trivial in the setting (Section \ref{sec:optCTRL}).
    \item the validation of the performance of the controller by applying it to a high fidelity model, thus rendering our controller as an add-on to existing tools to guarantee safety (Section \ref{sec:simulation}).
\end{enumerate}
The remainder of the paper is organized as follows: Section \ref{sec:modeling} discusses the AWE model used for simulation purposes as well as the guidance strategy and baseline controller. Section \ref{sec:modelforsafety} introduces the simplifications, abstractions, and the new reference frame used to derive a low dimensional model of the AWE system suitable for safety-control synthesis, with the safety controller being derived and introduced in Section \ref{sec:HJ}. Finally, in Section \ref{sec:simulation} we validate the controller in its hybrid control setup using the high fidelity AWE model and discuss the effects on power generation with Section \ref{sec:conclusion} providing concluding remarks and directions of future research.

\section{Modeling}
\label{sec:modeling}
The modeling of AWE systems has been well studied and a variety of definitions for the equations of motion have been derived. Yet commonly AWE systems require a high number of states in order to accurately capture both the dynamics of the kite or aircraft as well as the tether and the associated winch controller. Since the goal of this paper is to derive a safety-critical controller for the AWE system using HJ reachability analysis, we seek the development of a simplified yet accurate enough, from a control point of view, model.

In the following section, we begin by modeling the AWE system based predominantly on the work of \cite{Rapp2021} with extensions taken from \cite{Malz2019, Rapp2018} and \cite{Fechner2015}. This model will be used for simulation purposes in Section \ref{sec:simulation} and is presented to provide the necessary notation and understanding needed for the abstractions and synthesis, the primary contributions of this work, in Sections \ref{sec:modelforsafety} and \ref{sec:HJ}.

\subsection{Reference Frames} \label{subsec:referenceframe}
For the discussion of the dynamics of AWE systems, it is useful to introduce a variety of reference frames. The utilized reference frames (with the exception of the $\Gamma$ frame, Section \ref{subsec:gammaframe}) are common for AWE systems and have been extensively discussed in \cite{Rapp2021} and \cite{Jehle2014}. For the transformation from one reference frame to another, we introduce the transformation matrices $\mathbf{M}_{(\cdot) (\cdot)}$, where the first subscript indicates the  destination frame and the second subscript indicates the origin frame. Thus, as an example, the kinematic velocity in the $\tau$ frame, where $\tau$ denotes one of the employed frames, can be obtained from the body fix frame (B) by $(\mathbf{v}_{\mathrm{k}})_{\tau} = \mathbf{M}_{\tau \mathrm{B}} (\mathbf{v}_{\mathrm{k}})_{\mathrm{B}}$.
The transformation matrices for the reference frames are listed in the appendix. We utilize bold notation for vectors and matrices and regular notation for scalars. Furthermore, we utilize subscripts to denote vector elements, i.e., $(v_{\mathrm{k}})_{\tau, \mathrm{y}}$ denotes the $\mathrm{y}$ component of the kinematic velocity in the $\tau$ frame.

\subsubsection{Wind Frame}
The wind frame (W) is a rotation of the commonly used North-East-Down (NED or O) frame. The x-axis of the W frame is aligned with the mean wind direction (denoted by $\xi$). We assume that the mean wind direction is such that the z-axis of the W frame points upwards and the y-axis forms a right-hand reference frame (Fig. \ref{fig:frame}). 
\subsubsection{Tangential Frame}
The tangential frame ($\tau$) is centered at the position of the aircraft, such that the z-axis points towards the origin of the NED frame. To this end, the x-axis points towards the north direction, while the y-axis completes a right-hand reference frame. Since the $\tau$ frame moves together with the aircraft, its location is determined using the longitude ($\lambda$) and latitude ($\phi$) as well as the distance to the origin ($h_{\tau}$). The tangential frame is shown in purple in Fig. \ref{fig:frame}.
\subsubsection{Aerodynamic Frame}
The aircraft's kinematic velocity is denoted by $\mathbf{v}_{\mathrm{k}}$ and is composed of the aerodynamic velocity $\mathbf{v}_{\mathrm{a}}$ as well as the wind velocity $\mathbf{v}_{\mathrm{w}}$, i.e., $\mathbf{v}_{\mathrm{k}} = \mathbf{v}_{\mathrm{a}} + \mathbf{v}_{\mathrm{w}}$. This relationship holds true regardless of reference frame. For a discussion of the aerodynamic velocity, we define the aerodynamic frame (A) such that the velocity vector $\mathbf{v}_{\mathrm{a}}$ is aligned with the x-axis. For the derivation with respect to the NED frame (shown in blue in Fig. \ref{fig:frame}), we first rotate the NED frame by the course angle $\chi_{\mathrm{a}}$ and path angle $\gamma_{\mathrm{a}}$, resulting in the intermediate frame $\overline{A}$ (shown in orange in Fig. \ref{fig:frame}), before tilting the frame by the bank angle $\mu_{\mathrm{a}}$. The A frame is shown in green in Fig. \ref{fig:frame}.
\subsubsection{Body-Fixed Frame}
\begin{figure}
 	\centering
         \includegraphics[width=0.54\textwidth]{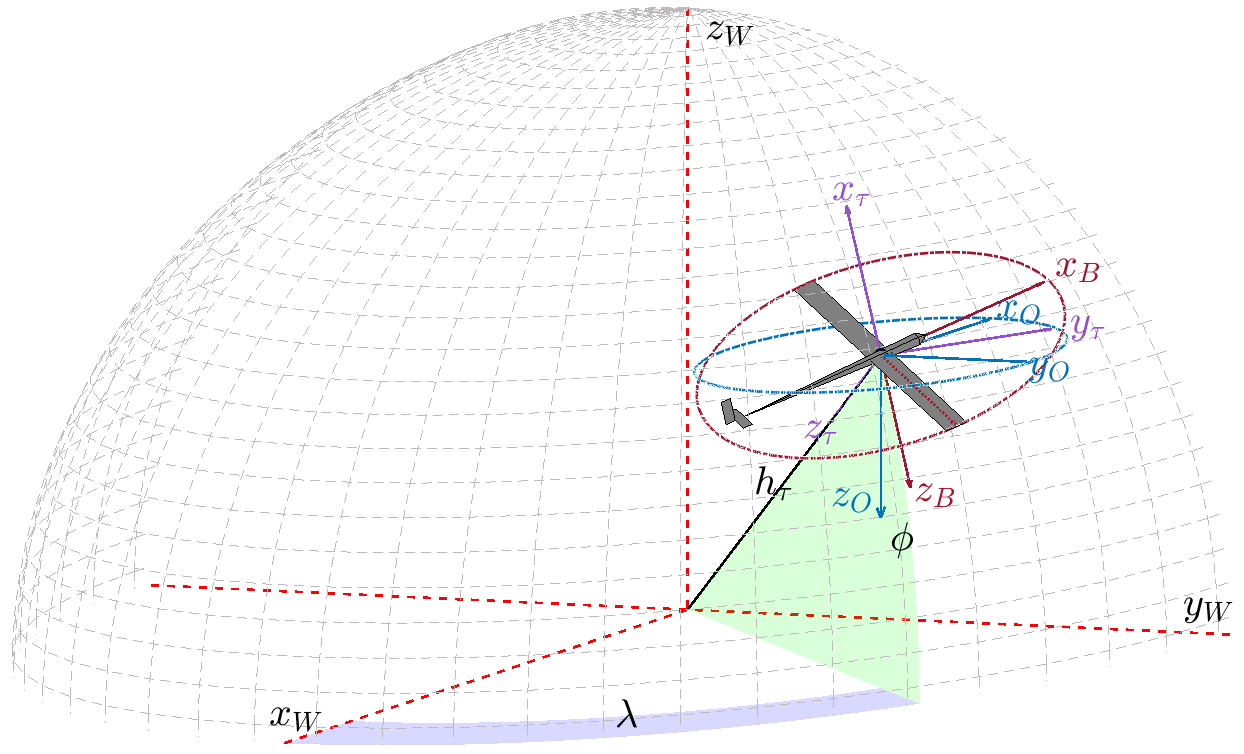}
	\includegraphics[width=0.45\textwidth]{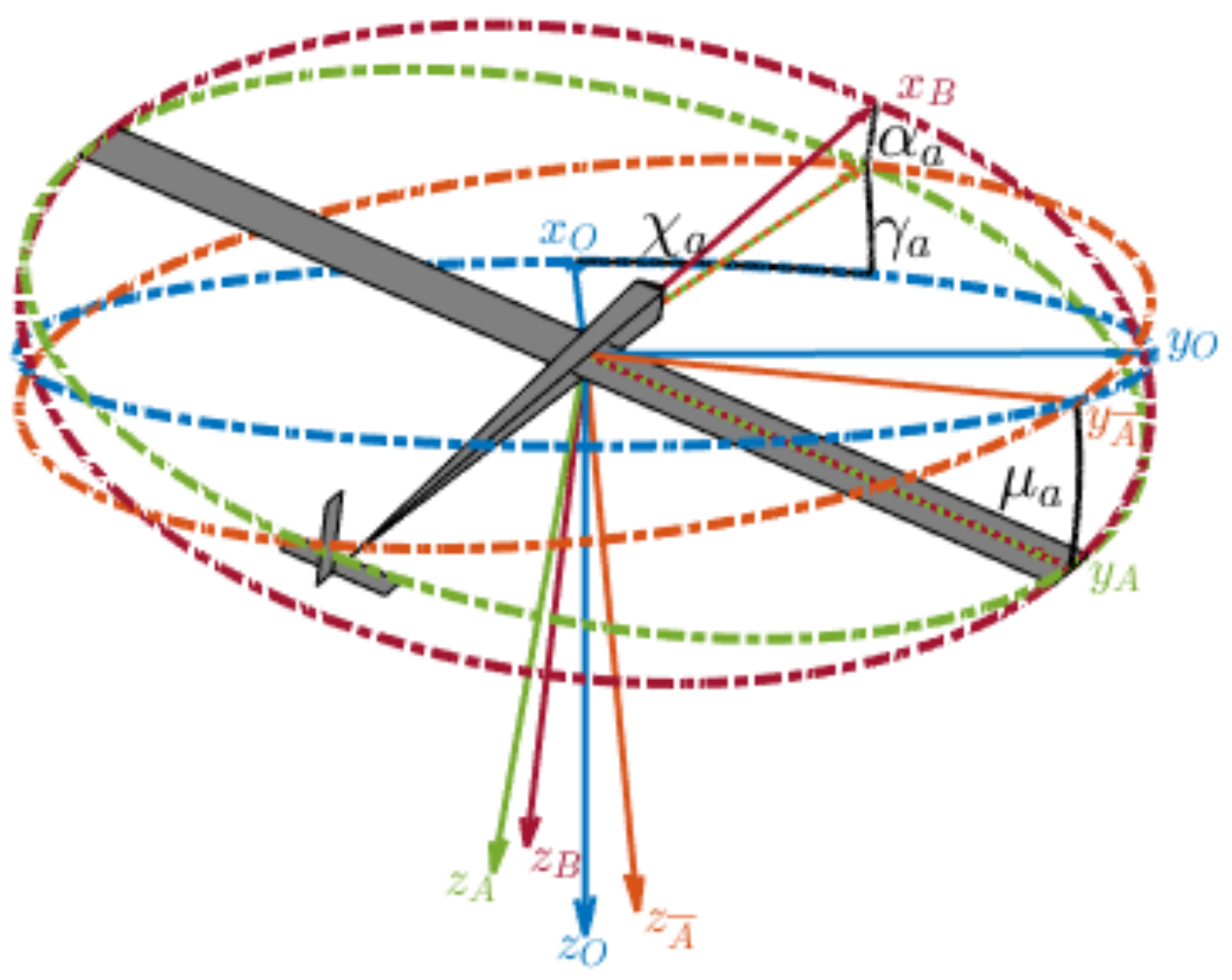}
	\caption{Visualization of the NED frame (blue), intermediate aerodynamic frame $\overline{A}$ (orange), the aerodynamic frame A (green), the body frame B (red) and the tangential frame (purple).}
	\label{fig:frame}
\end{figure}

The body-fixed frame (B) is used to calculate the aerodynamic forces acting on the aircraft. It can be obtained from the aerodynamic frame (shown in green in Fig. \ref{fig:frame}) using the angle of attack $\alpha_{\mathrm{a}}$ as well as the sideslip angle $\beta_{\mathrm{a}}$ (the sideslip angle is neglected in this work). The B frame is shown in red in Fig. \ref{fig:frame}.
 
\subsection{Aircraft Equations of Motion and Ground Station Model}
\label{subsec:equationsmotion}
For the AWE system considered in this paper, we simplify the modeling of the aircraft by neglecting the yaw, pitch, and roll rates as well as the sideslip angle $\beta_a$. As control inputs we consider the bank angle $\mu_a$ as well as the angle of attack $\alpha_{\mathrm{a}}$. A more complex model of the AWE system is outside of the scope of this paper. 

The position of the aircraft is defined in the W frame by $[\lambda, \phi, h_{\tau}]$ and it moves with velocity $v_a$ along the x-axis of the $\overline{A}$ frame. To connect the $\overline{A}$ frame with the NED frame, we add the course angle $\chi_a$ and path angle $\gamma_a$ as additional states. To influence the heading of the aircraft, we utilize the bank angle $\mu_{\mathrm{a}}$ and angle of attack $\alpha_{\mathrm{a}}$ as inputs to the system and assume that the sideslip angle $\beta_{\mathrm{a}}$ is negligible, i.e., $\beta_{\mathrm{a}} = 0$.

Therefore our initial set of states used are $[\lambda, \phi, h_{\tau}, v_{\mathrm{a}}, \chi_{\mathrm{a}}, \gamma_{\mathrm{a}}] \in \mathbb{R}^6$. The position propagation can be calculated from the kinematic velocity in the $\tau$ frame as follows:
\begin{align}
\dot{\lambda} & = \frac{(v_{\mathrm{k}})_{\tau, \mathrm{y}}}{\cos(\phi) h_{\tau}} \\
\dot{\phi} & = \frac{(v_{\mathrm{k}})_{\tau, \mathrm{x}}}{ h_{\tau}} \\
\dot{h}_{\tau} & = -(v_{\mathrm{k}})_{\tau, \mathrm{z}}
\end{align}
where $(v_{\mathrm{k}})_{\tau, \mathrm{x}}$, $(v_{\mathrm{k}})_{\tau, \mathrm{y}}$ and $(v_{\mathrm{k}})_{\tau, \mathrm{z}}$ denote the x, y and z components of the kinematic velocity in the $\tau$ frame, respectively. $(\mathbf{v}_{\mathrm{k}})_{\tau}$ is calculated by transforming the kinematic velocity from the O frame to the W frame, and then to the $\tau$ frame using the appropriate transformation matrix, $\mathbf{M}_{\mathrm{W} \mathrm{O}}$ and $\mathbf{M}_{\tau \mathrm{W}}$, respectively, found in the appendix. $(\mathbf{v}_{\mathrm{k}})_{\mathrm{O}}$ is derived from the aerodynamic and wind velocity
\begin{equation}
    (\mathbf{v}_{\mathrm{k}})_{\mathrm{O}} = \mathbf{M}_{\mathrm{O}\overline{\mathrm{A}}} \begin{bmatrix}
     v_{\mathrm{a}} , \: 0 , \: 0
    \end{bmatrix}_{\mathrlap{\overline{\mathrm{A}}}}^T + (\mathbf{v}_{\mathrm{W}})_{\mathrm{O}}
    \label{eqn:kinematicverlocity}
\end{equation}

By assuming the wind field stays stationary, we are able to calculate the derivative of the kinematic velocity using the gravitational force ($\mathbf{F}_{\mathrm{g}}$), aerodynamic force ($\mathbf{F}_{\mathrm{a}}$), as well as the force the tether exerts on the aircraft ($\mathbf{F}_{\mathrm{t}}$). For a complete derivation we refer to \cite{Rapp2021}.
\begin{equation}
\begin{bmatrix}
    \dot{v}_{\mathrm{a}} \\ \dot{\chi}_{\mathrm{a}} \\ \dot{\gamma}_{\mathrm{a}}
\end{bmatrix} = \frac{1}{m_{\mathrm{a}}} \begin{bmatrix}
1 & 0 & 0 \\
0 & \frac{1}{v_{\mathrm{a}} \cos{\gamma_{\mathrm{a}}}} & 0 \\
0 & 0 & -\frac{1}{v_{\mathrm{a}}}
\end{bmatrix} \Big( \mathbf{M}_{\overline{\mathrm{A}} \mathrm{O}} \big[ (\mathbf{F}_{\mathrm{g}})_{\mathrm{O}} + (\mathbf{F}_{\mathrm{t}})_{\mathrm{O}}\big] + \mathbf{M}_{\overline{\mathrm{A}} \mathrm{A}} \mathbf{M}_{\mathrm{A} \mathrm{B}} (\mathbf{F}_{\mathrm{a}})_{\mathrm{B}} \Big)
\end{equation}
where $m_{\mathrm{a}}$ denotes the mass of the aircraft. Note that the matrix $\mathbf{M}_{\overline{\mathrm{A}} \mathrm{O}} $ depends on the course and path angles $\chi_{\mathrm{a}}$ and $\gamma_{\mathrm{a}}$, $\mathbf{M}_{\overline{\mathrm{A}} \mathrm{A}} $ depends on the bank angle $\mu_{\mathrm{a}}$ and $\mathbf{M}_{\mathrm{A} \mathrm{B}} $ depends on the angle of attack $\alpha_{\mathrm{a}}$ and sideslip angle $\beta_{\mathrm{a}}$.

\subsubsection{Ground Station Model}
The ground station serves as the power generator for the AWE system. The winch controller within the ground station is responsible for reeling out and reeling in the tether during flight. The winch is essentially a drum on which the tether is wound. The position of the drum is defined by $\theta_{\mathrm{w}}$ and the reeled out tether length is simply calculated as $l_{\mathrm{tether}} = r_{\mathrm{w}} \theta_{\mathrm{w}}$, where $r_{\mathrm{w}}$ is the drum radius. Using the same dynamics as in \cite{Rapp2018}, we can thus describe the winch using a second order system.
\begin{align}
    \dot{\theta}_{\mathrm{w}} & = \omega_{\mathrm{w}} \\
    \dot{\omega}_{\mathrm{w}} & = \frac{1}{J_{\mathrm{w}}} \big(
    r_{\mathrm{w}} || \mathbf{F}_{\mathrm{w}} ||_2 - \nu_{\mathrm{w}} \dot{\theta}_{\mathrm{w}} + M_{\mathrm{c}}
    \big)
    \label{eqn:grounddynamcis}
\end{align}
where $\nu_{\mathrm{w}}$ is the friction coefficient, $\omega_{\mathrm{w}}$ is the rotation rate of the drum, $\mathbf{F}_{\mathrm{w}}$ is the tether force at the ground station (equal to $\mathbf{F}_{s,0}$ in \ref{subsec:tetherforce}), and $J_{\mathrm{w}}$ is the inertia of the winch. For the control of the winch in the simulation study discussed in Section \ref{sec:simulation}, we regulate the control moment $M_{\mathrm{c}}$ using a PI-controller that attempts to track a given reference tether force, $F_{\mathrm{ref}}$. 

\subsubsection{Gravitational and Aerodynamic Forces}
The gravitational force can simply be expressed in the O frame as 
\begin{equation}
(\mathbf{F}_{\mathrm{g}})_{\mathrm{O}} = \begin{bmatrix}
0 , \: 0 , \: m_{\mathrm{a}} g \end{bmatrix}^T \in \mathbb{R}^3
\end{equation}
where $g$ is the gravitational constant. As mentioned earlier, we assume yaw, pitch and roll rates to be zero and neglect the effects of the sideslip angle $\beta_{\mathrm{a}}$. This simplifies the aerodynamic coefficients derived in \cite{Malz2019}, to
\begin{equation}
\begin{bmatrix}
C_{\mathrm{x}}\\ C_{\mathrm{y}} \\ C_{\mathrm{z}}\end{bmatrix}= \begin{bmatrix}
C_{\mathrm{x}, 0}(\alpha_{\mathrm{a}}) + C_{\mathrm{x}, \delta_{\mathrm{e}}}(\alpha_{\mathrm{a}})\delta_{\mathrm{e}} \\
 C_{\mathrm{y}, \delta_{\mathrm{a}}}(\alpha_{\mathrm{a}})\delta_{\mathrm{a}} + C_{\mathrm{y}, \delta_{\mathrm{r}}}(\alpha_{\mathrm{a}})\delta_{\mathrm{r}}\\
C_{\mathrm{z}, 0}(\alpha_{\mathrm{a}}) + C_{\mathrm{z}, \delta_{\mathrm{e}}}(\alpha_{\mathrm{a}})\delta_{\mathrm{e}}
\end{bmatrix}
\end{equation}
where we consider the aileron ($\delta_{\mathrm{a}}$), elevator ($\delta_{\mathrm{e}}$) and rudder ($\delta_{\mathrm{r}}$) deflections to be constant and all $\alpha_{\mathrm{a}}$ dependent coefficients are approximated using second order polynomials. Using the aerodynamic coefficients we can then calculate the aerodynamic force of the aircraft as
\begin{equation}
(\mathbf{F}_{\mathrm{a}})_{\mathrm{B}} = \frac{1}{2} \rho \mathrm{S}_{\mathrm{ref}} v_{\mathrm{a}}^2 \begin{bmatrix}
C_{\mathrm{x}} , \: C_{\mathrm{y}} , \: C_{\mathrm{z}}\end{bmatrix}^T
\end{equation}
where $\mathrm{S}_{\mathrm{ref}}$ is the aerodynamic reference area corresponding to the projected surface area of the aircraft wing \cite{Malz2019}, and $\rho$ is the air density.

\subsubsection{Tether Forces}
\label{subsec:tetherforce}
For the calculation of the tether forces ($\mathbf{F}_{\mathrm{t}}$) that act on the aircraft, we use models of varying complexity. For the controllers, we rely on a straight tether approximation, whereas for simulation purposes, we use a tether model similar to the model derived in \cite{Fechner2015}. We begin by deriving the most complex of the three tether models employed in this paper.

\begin{figure}
 	\centering
	\includegraphics[width=0.6\textwidth]{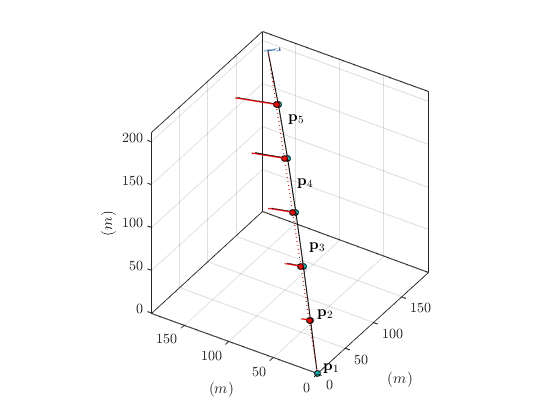}
	\caption{Visualization of the tether modeled as a straight tether (red) and as a fixed number of lumped masses (blue). The velocities of the point masses are shown by the red and blue vectors respectively.}
	\label{fig:tether}
\end{figure}

Let us first consider a fixed number of lumped masses connected by $n$ spring-damper elements. The length of each segment is denoted by $l_s$ and the spring and damping constants for each segment are denoted by $k$ and $c$, respectively. Each tether segment is modeled as a point mass with position $\mathbf{p}_{i}$ and velocity $\mathbf{v}_{i}$, as shown in Fig. \ref{fig:tether}. 
The equations of motion of the tether segment are given by
\begin{align}
    \dot{\mathbf{p}}_{i} & = \mathbf{v}_{i} \\
    \dot{\mathbf{v}}_{i} & = m_{t} \mathbf{F}_{i}
\end{align}
where $m_{t}$ is the mass of an individual segment and $\mathbf{F}_{i}$ is the tether segment force. The tether segment force for segment i is given by
\begin{equation}
    \mathbf{F}_{i} = \mathbf{F}_{s,i+1} - \mathbf{F}_{s,i} - \mathbf{F}_{g, s} + \mathbf{F}_{a,i}
\end{equation}
where $\mathbf{F}_{g, s}$ is the gravitational force acting on the tether segment, $\mathbf{F}_{a,i}$ is the aerodynamic drag acting on the i-th segment, and $\mathbf{F}_{s,i}$ is the tensile force of the i-th segment.

The tensile force is calculated according to Hooke's law
\begin{equation}
    \mathbf{F}_{s,i} = \Big(k \big(|| \mathbf{s}_{i}||_2 -l_s\big) + c \big( \frac{\mathbf{s}_{i}}{|| \mathbf{s}_{i}||_2} \mathbf{s}_{v,i} \big) \Big) \frac{\mathbf{s}_{i}}{|| \mathbf{s}_{i}||_2}
    \label{eqn:tensile-force}
\end{equation}
where $\mathbf{s}_{i} = \mathbf{p}_{i} - \mathbf{p}_{i-1}$ and $\mathbf{s}_{v,i} = \mathbf{v}_{i} - \mathbf{v}_{i-1}$. Thus the maximum tether force is exerted on the final segment attached to the aircraft, i.e., $\mathbf{F}_{t} = \mathbf{F}_{s,n+1}$, with $\mathbf{s}_{n+1} = \mathbf{p}_{\mathrm{aircraft}} - \mathbf{p}_{n}$ and $\mathbf{s}_{v,n+1} = \mathbf{v}_{\mathrm{aircraft}} - \mathbf{v}_{n}$ and $\mathbf{p}_{\mathrm{aircraft}}$ and $\mathbf{v}_{\mathrm{aircraft}}$ are the position and velocity, respectively, of the aircraft in Cartesian coordinates.

For the tether drag calculation, we begin by introducing the apparent air velocity, $\mathbf{v}_{a,i}$, composed of the wind speed at the height of the i-th particle ($\mathbf{v}_{\mathrm{w},i}$), as well as the average segment velocity
\begin{equation}
    \mathbf{v}_{a,i} = \mathbf{v}_{\mathrm{w},i} - \frac{\mathbf{v}_{i+1} + \mathbf{v}_{i}}{2}
\end{equation}
Each segment is modeled as a cylinder, thus the drag is caused predominantly by the velocity perpendicular to the tether segment
\begin{equation}
    \mathbf{v}_{a,i, \perp} = \mathbf{v}_{a,i} - \Big( \frac{\mathbf{s}_{i}^T \mathbf{v}_{a,i}}{||\mathbf{s}_{i}||_2} \Big) \frac{\mathbf{s}_{i}}{||\mathbf{s}_{i}||_2}
\end{equation}
Using the perpendicular velocity components, the final tether drag is given by 
\begin{equation}
    \mathbf{F}_{a,i} = \frac{1}{2} \rho C_{d, t} \mathbf{v}_{a,i, \perp} ||\mathbf{v}_{a,i, \perp}||_2 A_{\mathrm{eff}, t}
\end{equation}
where $C_{d, t}$ is the tether drag coefficient and $A_{\mathrm{eff}, t}$ is the projected tether area perpendicular to $\mathbf{v}_{a,i}$,
\begin{equation}
    A_{\mathrm{eff}, t} = d_{t} || \mathbf{s}_{i} - \mathbf{s}_{i} \frac{\mathbf{v}_{a,i}}{||\mathbf{v}_{a,i}||_2}||_2
\end{equation}
where $d_{t}$ is the tether diameter.

This complex tether model requires $6n$ additional states, where $n$ is the number of segments used, and is therefore only suited for simulation purposes. For the controller synthesis we use two simplified tether models that rely on a straight tether approximation. Both models will be introduced in Section \ref{subsec:simplifiedtether} and Section \ref{subsec:baselinecontroller}.

\subsubsection{Wind Field Model}
To capture the varying wind speeds at different altitudes we employ the wind shear model provided by the MATLAB Aerospace Toolbox \cite{AerospaceToolbox2021a}.
\begin{equation}
    (\mathbf{v}_{\mathrm{\mathrm{shear}}})_{\mathrm{O}} = W_{20} \frac{\ln(\frac{h}{z_0})}{\ln(\frac{20}{z_0})}
\end{equation}
where $h$ is the altitude of the aircraft in feet, $z_0$ is a constant equal to 0.15 feet, and $W_{20}$ is the measured wind speed at an altitude of 20 feet.

In addition to the wind shear, we also model atmospheric turbulence in form of a continuous Dryden wind turbulence model. The Dryden model is a stochastic gust model, whereby the linear and angular velocities of the atmospheric turbulence are modeled as spatially varying stochastic processes, each with a specific power spectral density. The longitudinal, lateral, and vertical component spectra functions are provided by Military Handbook MIL-HDBK-1797B \cite{MilitaryHandbook}. For the implementation of the Dryden model, we utilize the continuous Dryden model block provided by the MATLAB Aerospace Toolbox \cite{AerospaceToolbox2021a}.

Neglecting the angular rates of the atmospheric turbulence and assuming the pitch, roll and yaw rates of the aircraft stay constant, the wind turbulence is given by $(\mathbf{v}_{\mathrm{\mathrm{turb}}})_{\mathrm{W}} = [u_{g}, v_{g}, w_{g}]^T \in \mathbb{R}^3$, where $u_{g}$ is the longitudinal turbulence velocity aligned along the horizontal relative mean wind vector, $v_{g}$ is the lateral turbulence velocity and $w_{g}$ is the vertical turbulence velocity. Due to the varying altitudes of the aircraft, the terms $u_{g}$, $v_{g}$ and $w_{g}$ are computed by passing a band-limited white noise signal through two sets of forming filters, one for low altitudes and one for high altitudes, and then interpolating the results.

Finally the wind velocity acting on the aircraft is computed by adding the turbulence to the wind shear $(\mathbf{v}_{\mathrm{\mathrm{W}}})_{\mathrm{O}} = (\mathbf{v}_{\mathrm{\mathrm{turb}}})_{\mathrm{O}} + (\mathbf{v}_{\mathrm{\mathrm{shear}}})_{\mathrm{O}}$.

\subsection{Optimal Flight Path} \label{subsec:optimalflightpath}
The fundamental idea behind Ground-Gen AWE is to transfer the force from the aerodynamic lift of the aircraft or kite to the connected tether \cite{Diehl2013}. Similar to the blades of a wind turbine, this can be done by moving the aircraft perpendicular to the mean wind direction. To this end, it has become common practice in the AWE community to adopt a figure eight flight pattern. We thus use a Lissajous curve, $\Gamma$, lying on a sphere as the reference flight path. The two-dimensional curve can be described by its longitude $\lambda_\Gamma$ and latitude $\phi_\Gamma$. We parameterize $\Gamma$ on $\mathbb{S}^2$ using the arc length $s$. In Cartesian coordinates the curve is then given by
\begin{equation}
    \Gamma_{\mathrm{P}}(s) = \begin{bmatrix}\cos{\lambda_\Gamma(s)} \cos{\phi_\Gamma(s)} \\
    \sin{\lambda_\Gamma(s)} \cos{\phi_\Gamma(s)} \\
    \sin{\phi_\Gamma(s)}
    \end{bmatrix} h_{\tau}
\end{equation}
For the specific Lissajous curve, we opt for the Lemniscate of Booth \cite{Rapp2019}, a commonly used figure eight curve, using the height/width ratio of $a_{\mathrm{Booth}}/b_{\mathrm{Booth}} = 120/200$. The longitude and latitude of the Lemniscate of Booth can be calculated as follows:
\begin{align}
    \lambda_\Gamma(s) & = \frac{1}{h_{\tau}} \frac{b_{\mathrm{Booth}} \sin{s}}{1 + \big( \frac{a_{\mathrm{Booth}}}{b_{\mathrm{Booth}}}\big)^2 \cos^2{s}} \\
    \phi_\Gamma(s) & = \frac{1}{h_{\tau}} \frac{a_{\mathrm{Booth}} \sin{s} \cos{s}}{1 + \big( \frac{a_{\mathrm{Booth}}}{b_{\mathrm{Booth}}}\big)^2 \cos^2{s}}
\end{align}
We define our reference curve to lie centered on the horizon. However, during flight, we rotate the curve by $\psi_0$ to assist with power generation. Thus the tracking curve used during flight is given by 
\begin{equation}
    \Gamma(s, \psi_0) = \underbrace{\begin{bmatrix}
    \cos(\psi_0) & 0 &-\sin(\psi_0) \\
    0 & 1 & 0 \\
    \sin(\psi_0) & 0 & \cos(\psi_0)
    \end{bmatrix}}_{\mathbf{M}_{\mathrm{W}\mathrm{P}}} \Gamma_{\mathrm{P}}(s)
\end{equation}
Finally, the tangent of the curve is given by $\mathbf{t}(s) = \frac{d \Gamma}{d s} = \frac{\partial \Gamma}{\partial \lambda_\Gamma} \frac{\partial \lambda_\Gamma}{\partial s}+ \frac{\partial \Gamma}{\partial \phi_\Gamma} \frac{\partial \phi_\Gamma}{\partial s}$.

\subsection{Guidance Strategy}
\label{subsec:guidancestrategy}
Since we only derive the safety controller for the traction phase, the most critical part of flight (neglecting take-off and landing), we refer to \cite{Rapp2021} for a complete discussion of the guidance strategy employed during retraction and transition from retraction to traction.

As shown in Fig. \ref{fig:gamma}, the shortest path from the position of the aircraft (denoted by $\mathbf{K}$) to the closet point on the tracking curve (denoted by $\mathbf{C}$), is given by the geodesic vector $\boldsymbol\sigma$. For perfect path tracking, we need to minimize $||\boldsymbol\sigma||_2$. Taking the shortest path, however, would result in the aircraft intercepting the curve perpendicularly. In practice, this is not desirable and thus we need to introduce a commanded flight direction that results in the distance to the tracking curve being minimized, while also ensuring that when the aircraft intercepts the tracking curve, its kinematic velocity is aligned with the tangent of the curve, $\mathbf{t}$.

The commanded flight path can be defined on $\mathbb{S}^2$, independently of the distance to the origin by simply scaling the parameters $a_{\mathrm{Booth}}$ and $b_{\mathrm{Booth}}$ by $1/h_{\tau}$. The optimal course angle $(\chi)_{\tau}$ should point the aircraft along the geodesic when far from the curve, however, point perpendicular to the geodesic to align with $\mathbf{t}(s)$ when close to the curve. We set the optimal flight path angle to zero to prevent the aircraft from going into a nose dive in order to quickly gain speed and reach the curve, something that would be counterintuitive to energy production. 

\begin{figure}
 	\centering
	\includegraphics[width=0.3\textwidth]{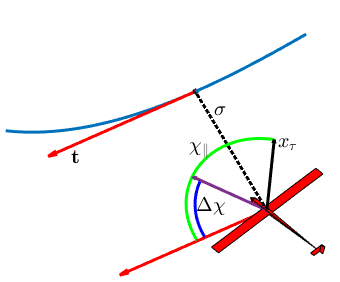}
	\caption{Visualization of the guidance strategy to produce the optimal course heading (purple arrow).}
	\label{fig:cmd}
\end{figure}

To this end, let us define the optimal course and path angles as 
\begin{align}
    (\gamma_{\mathrm{cmd}})_{\tau} & = 0 \\
    (\chi_{\mathrm{cmd}})_{\tau} & = (\chi_{\parallel})_{\tau} + (\Delta\chi)_{\tau}
\end{align}
with $(\chi_{\parallel})_{\tau} = \arctan{\frac{t_{y}}{t_{x}}}$, where $t_{x}$ and $t_{y}$ denote the x and y components of the curve tangent. The second component, $\Delta\chi_{\tau}$, of the command course angle ensures that as we approach the tracking curve, we align ourselves with the curve tangent $\mathbf{t}$. This is done by introducing a tuning parameter $\delta_0$, allowing us to calculate course change as
\begin{equation}
    (\Delta\chi)_{\tau} = \arctan{\frac{\mathrm{sign}(\sigma) ||\boldsymbol\sigma||_2}{\delta_0}}
\end{equation}
This tracking strategy is similar to that of \cite{Jehle2014}, and is visualized in Fig. \ref{fig:cmd}.

\subsection{Benchmark Controller}
\label{subsec:baselinecontroller}
Since the safety controller derived in this paper considers worst-case environmental conditions, to alleviate conservatism we will activate it only when safety is at stake. Furthermore, the safety controller is only configured for the flight of the aircraft in the traction phase. To this end, we employ a primary controller synthesized using nonlinear dynamic inversion (NDI), an approach common in the aerospace sector \cite{Smith1998}. For the NDI controller, we begin by calculating the optimal course and path angle rates ($\dot{\chi}_{\mathrm{cmd}}$ and $\dot{\gamma}_{\mathrm{cmd}}$) that lead to perfect curve tracking. Together with the course and path angle errors, we are able to compute the necessary pseudo-control inputs
\begin{align}
    \nu_{\chi} & = \dot{\chi}_{\mathrm{cmd}} + k_{p, \chi} (\chi_{\mathrm{cmd}} - \chi) - \dot{\chi}_{\mathrm{est}}\\
    \nu_{\gamma} & = \dot{\gamma}_{\mathrm{cmd}} + k_{p, \gamma} (\gamma_{\mathrm{cmd}} - \gamma) - \dot{\gamma}_{\mathrm{est}}
\end{align}
where $k_{p, \chi}$ and $k_{p, \gamma}$ are control gains. Furthermore, $\dot{\chi}_{\mathrm{est}}$ and $\dot{\gamma}_{\mathrm{est}}$ are the estimated path angle rates based on a simplified version of the aircraft dynamics using a straight tether approximation, $\widehat{\mathbf{F}}_{t}$. Using the pseudo control inputs, we can construct the optimal control input, $u_{\mathrm{NDI}}$, using the inverted aircraft dynamics. For detailed controller synthesis of the NDI controller, we refer to \cite{Rapp2021}. 

However, unlike in \cite{Rapp2021}, we cannot assume perfect tension tracking. When the safety controller is activated, the tether tension will be reduced. We, therefore, update the simplified model of the NDI controller to use the true tether force, $||\mathbf{F}_{t}||_2$, projected along a straight line towards the origin. Thus the modified straight tether approximation is given by $\widehat{\mathbf{F}}_{t} = \frac{\mathbf{p}_{\mathrm{aircraft}}}{||\mathbf{p}_{\mathrm{aircraft}}||_2} ||\mathbf{F}_{t}||_2$, where $\mathbf{p}_{\mathrm{aircraft}}$ is the position of the aircraft, and $\mathbf{F}_{t}$ is given by $\mathbf{F}_{s,n+1}$ as in \eqref{eqn:tensile-force}.

\section{Model Abstraction and Safety Considerations}
\label{sec:modelforsafety}
For the derivation of the safety controller presented in Section \ref{sec:HJ}, the previously introduced model does not suffice, since it requires too many states for synthesis. We thus begin by introducing a simplification of the tether model that allows us to introduce adversarial winch control into the AWE dynamics. Together with a new reference frame and the subsequent safety control model, this section provides the first of our four key contributions.
\subsection{Simplified Tether Model}
\label{subsec:simplifiedtether}
As mentioned in the previous sections, a full particle tether model requires too many states, prohibiting its adoption for the safety controller. From field tests conducted with an aircraft in \cite{Jehle2014}, for a small aircraft the tether forces far exceed the gravitational forces and a straight tether approximation is reasonable. However, if larger systems with larger tethers are employed, the tether sag will need to be taken into account. For the AWE system considered in this paper, for the purpose of controller synthesis, we assume that the tether sag remains negligible and, therefore, opt to approximate the position and velocity of the tether segments. To this end, we assume that the position and velocity of the tether segments are distributed evenly along a straight line between the origin and the position of the aircraft. Thus the tether particle's position and velocity can be computed as the x, y, and z projections of the straight tether up to the i-th particle, i.e.,
\begin{align}
    \mathbf{p}_{i} & = \frac{i h_{\tau}}{n+1} \begin{bmatrix}
    \cos(\phi) \cos(\lambda) \\
    \cos(\phi) \sin(\lambda) \\
    \sin(\phi)
    \end{bmatrix}\\
    \mathbf{v}_{i} & = \frac{i \dot{h}_{\tau}}{n+1} \begin{bmatrix}
    \cos(\phi) \cos(\lambda) \\
    \cos(\phi) \sin(\lambda) \\
    \sin(\phi)
    \end{bmatrix} 
    -\frac{i h_{\tau}}{n+1} \begin{bmatrix}
    \sin(\phi) \dot{\phi} \cos(\lambda) + \cos(\phi) \sin(\lambda) \dot{\lambda}\\
    \sin(\phi) \dot{\phi} \sin(\lambda) - \cos(\phi) \cos(\lambda) \dot{\lambda}\\
    -\cos(\phi) \dot{\phi}
    \end{bmatrix}
    \label{eqn:straighttether_particles}
\end{align}
In Fig. \ref{fig:tether} a comparison of the straight tether and the full tether model is shown. Recall that the tether force acting on the aircraft is given by the spring force of the (n+1)-th segment, i.e., $\mathbf{F}_{t} = \mathbf{F}_{s,n+1}$. The dominating factor in Hooke's law is given by the term $||\mathbf{s}_{n+1}||_2 -l_s$, which is the difference between the segment length and the distance between the final tether point mass and the aircraft. Having $||\mathbf{s}_{n+1}||_2 -l_s > 0$ implies the distance is greater than the length of the tether and thus the tether is under tension. Modeling this term accurately is imperative for calculating the tether force, as the slightest deviations result in vastly inaccurate force calculations. To this end, let us introduce $\Delta_{t} = || \mathbf{s}_{n+1}||_2 -l_s$ as an additional state. By \eqref{eqn:tensile-force}, the tether force acting on the aircraft can be calculated as 
\begin{equation}
    \mathbf{F}_{t} = \Big(k \Delta_{t} + c \big( \frac{\mathbf{s}_{n+1}}{|| \mathbf{s}_{n+1}||_2} \mathbf{s}_{v,{n+1}} \big) \Big) \frac{\mathbf{s}_{n+1}}{|| \mathbf{s}_{n+1}||_2}
    \label{eqn:tetherforce}
\end{equation}
where the position and velocity of the tether point masses is computed using \eqref{eqn:straighttether_particles}.

The derivative of $\Delta_{t}$ is then given by 
\begin{equation}
    \dot{\Delta}_{t} =  \frac{\mathbf{s}_{v,n+1}^T \mathbf{s}_{n+1}}{||\mathbf{s}_{n+1}||_2} - \frac{r_{\mathrm{w}} \dot{\theta}_{\mathrm{w}}}{n+1}
\end{equation}
where $\dot{l}_s$, the reel-out speed of the tether, is substituted according to the winch dynamics \eqref{eqn:grounddynamcis}. In order to keep the number of system states minimal, we need to decouple the winch system from the aircraft dynamics. To this end, we model the effects of the winch as a scalar disturbance, $d_{\Delta_{t}} \coloneqq \dot{\Delta}_{t}$.
By accounting for worst-case winch behavior, the system can be considered robust to arbitrary winch control. To compute the expected range of values that the derivative of $\Delta_{t}$ might take, we run a simulation of the AWE simulation using the baseline controller, introduced previously. 

\begin{figure}[ht]
 	\centering
	\includegraphics[width=0.6\textwidth]{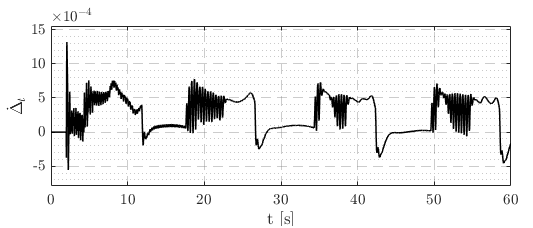}
	\caption{$\dot{\Delta}_{t}$ during the traction phase of flight, using the baseline NDI controller.}
	\label{fig:tether_diff_dot}
\end{figure}

In Fig. \ref{fig:tether_diff_dot}, $\dot{\Delta}_{t}$ for the traction phase of flight is shown. Since $\dot{\Delta}_{t}$ never exceeds $0.0015 \frac{m}{s}$, we constrain the disturbance, $d_{\Delta_{t}}$, to lie within the interval $[-d_{\Delta_{t},\max}, d_{\Delta_{t},\max}]$ with $d_{\Delta_{t},\max} = 0.005 \frac{m}{s}$.

\subsection{Adversarial Wind Turbulence}
\label{subsec:adverserialwindturbulence}
Since we model the wind turbulence as a stochastic process, we need to ensure that the safety controller can account for worst-case wind gusts, i.e., wind gusts that drive the system away from its target trajectory, and/or lead to a tether rupture. To this end, we model the wind turbulence as an adversarial disturbance input to the system. Thus the wind velocity for controller synthesis is given by
\begin{equation}
    (\mathbf{v}_{\mathrm{\mathrm{W}}})_{\mathrm{O}} = (\mathbf{v}_{\mathrm{\mathrm{shear}}})_{\mathrm{O}} + (\mathbf{d}_{\mathrm{\mathrm{turb}}})_{\mathrm{O}}
    \label{eqn:windvelocity2}
\end{equation}
where $\mathbf{d}_{\mathrm{\mathrm{shear}}} \in \mathbb{R}^3$ is an additional disturbance vector

\begin{figure}[ht]
 	\centering
	\includegraphics[width=0.6\textwidth]{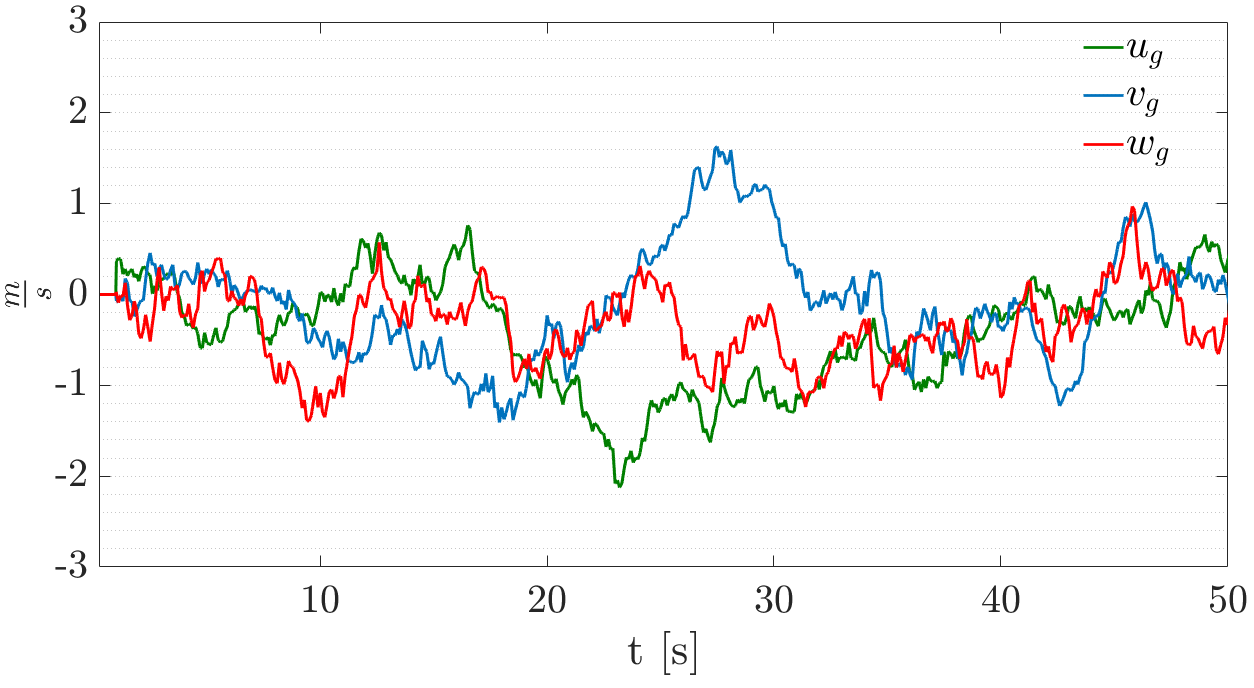}
	\caption{The behavior of the wind gusts during 50 seconds of flight.}
	\label{fig:turbulence_bounds}
\end{figure}

Similar to how we computed the bounds for $d_{\Delta_{t}}$, we simulate the flight of the aircraft using the baseline controller and analyze the behavior of $u_{g}$, $v_{g}$ and $w_{g}$. The wind gusts are shown in Fig. \ref{fig:turbulence_bounds}. From the behaviour $u_{g}$, $v_{g}$ and $w_{g}$, we can safely choose each element of $\mathbf{d}_{\mathrm{\mathrm{turb}}}$ to be bounded by $\pm 4 \frac{m}{s}$.

\subsection{$\Gamma$ Frame and Safety Control Model}
\label{subsec:gammaframe}
\begin{figure}[ht]
 	\centering
	\includegraphics[width=0.4\textwidth]{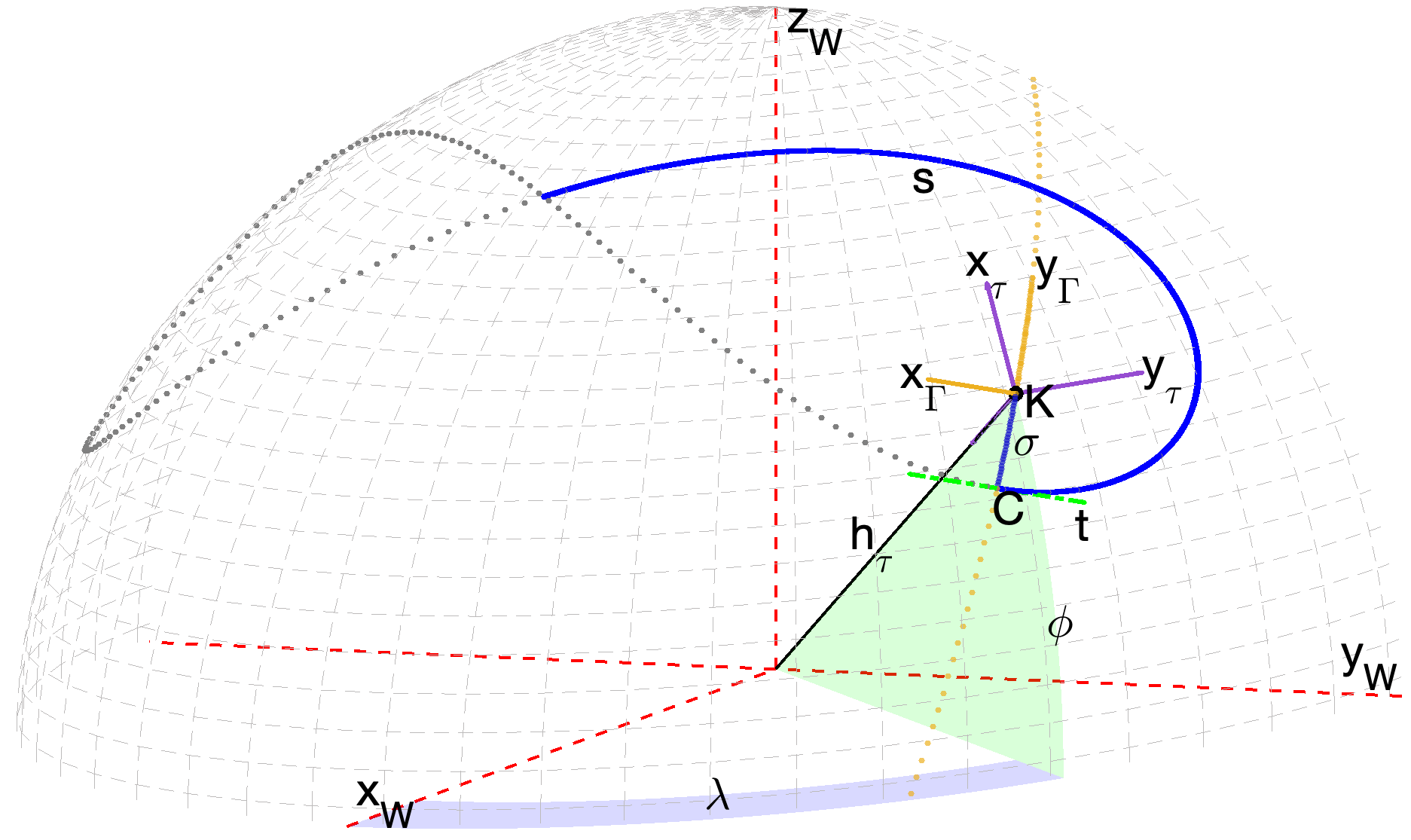}
	\caption{Visualization of the $\Gamma$ frame. $s$ is used to denote the position along the reference curve. The geodesic vector, $\sigma$, describes the distance to the reference curve.}
	\label{fig:gamma}
\end{figure}

    As visualized in Fig. \ref{fig:gamma}, the geodesic vector pointing from $\mathbf{C}$ along the geodesic towards $\mathbf{K}$, is always orthogonal to $\mathbf{t}$. The derivative of $\Gamma$ together with the direction of the geodesic vector can, therefore, be used as basis vectors to construct another reference frame that will become useful during gridding, as discussed in Section \ref{sec:HJ}. This new reference frame will be referred to as the $\Gamma$ frame. Any point on $\mathbb{S}^2$ defined using the longitude $\lambda$ and latitude $\phi$ can also be defined using the $\Gamma$ frame, where the first coordinate, $s$, describes an arbitrary point along the curve $\Gamma$, and the second coordinate $\sigma = ||\boldsymbol\sigma||_2$ describes the geodesic distance in meters perpendicular to $\mathbf{t}(s)$. The calculation of the geodesic for a sphere is derived from \cite{OpreaJohn2007} and relies on solving the geodesic equations for a sphere.

The derivatives of $s$ can be obtained by projecting the kinematic velocity of the aircraft onto $\mathbf{t}$. Similarly, the derivative of $\sigma$ is obtained by projecting the kinematic velocity onto a perpendicular rotation of $\mathbf{t}$, denoted by $\mathbf{t}_{\perp}$.
\begin{align}
    \dot{s} & = \frac{(\mathbf{t})_{\tau} (\mathbf{v}_k)_{\tau}}{||\mathbf{t}||_2 l_\Gamma} \\
    \dot{\sigma} & = \frac{(\mathbf{t}_{\perp})_{\tau} (\mathbf{v}_k)_{\tau}}{||\mathbf{t}_{\perp}||_2}
\end{align}
where $l_\Gamma$ denotes the arc length of the Lissajous curve $\Gamma$.
Using the $\Gamma$ frame allows us to replace the longitude and latitude resulting in the final set of states used for controller synthesis
\begin{equation}
    \mathbf{x} \coloneqq [s, \sigma, h_{\tau}, v_{\mathrm{a}}, \chi_{\mathrm{a}}, \gamma_{\mathrm{a}}, \Delta_{t}]^T \in \mathbb{R}^7
    \label{eqn:states}
\end{equation}
The transformation between the NED frame and the $\Gamma$ frame is not bijective, thus we need to take the velocity vectors into account to derive a deterministic transformation. To this end, we choose the transformation from the NED frame to the $\Gamma$ frame, such that the error between $(\mathbf{v}_{\mathrm{k}})$ and $\mathbf{t}$ is minimized.
Finally, we can summarize the equations of motion for the simplified aircraft model as $\dot{\mathbf{x}} = \mathbf{f}(\mathbf{x},\mathbf{u},\mathbf{d})$.
The control inputs, $\mathbf{u}$, and disturbances, $\mathbf{d}$, are defined as
\begin{equation}
    \mathbf{u} \coloneqq \begin{bmatrix} \alpha_{\mathrm{a}} \\ \mu_a \end{bmatrix} \in U \subset \mathbb{R}^2, \quad \mathbf{d} \coloneqq \begin{bmatrix}
        d_{\Delta_{t}} \\ \mathbf{d}_{\mathrm{\mathrm{turb}}}
    \end{bmatrix} \in D \subset \mathbb{R}^4
    \label{eqn:ud}
\end{equation}
The aircraft dynamics are composed of $\hat{\mathbf{f}}$, affected only by the control inputs, and $\mathbf{f}_C(\mathbf{x},\mathbf{d})$, affected only by the disturbances and incorporating all remaining additive terms, i.e., $\mathbf{f}(\mathbf{x},\mathbf{u},\mathbf{d}) = \hat{\mathbf{f}}(\mathbf{x},\mathbf{u}) + \mathbf{f}_C(\mathbf{x},\mathbf{d})$.
Decomposing the dynamics serves both the presentation of the final system dynamics, as well as the derivation of the Hamiltonian in the subsequent section. Since the control inputs affect only the aircraft velocity, course, and path angles, with a slight abuse of notation, we omit the four remaining states derivatives that are zero, allowing us to summarize $\hat{\mathbf{f}}(\mathbf{x},\mathbf{u})$ as
\begin{equation}
    \hat{\mathbf{f}}(\mathbf{x},\mathbf{u}) = \frac{1}{m_{\mathrm{a}}} \begin{bmatrix}
1 & 0 & 0 \\
0 & \frac{1}{v_{\mathrm{a}} \cos{\gamma_{\mathrm{a}}}} & 0 \\
0 & 0 & -\frac{1}{v_{\mathrm{a}}}
\end{bmatrix} \mathbf{M}_{\overline{\mathrm{A}} \mathrm{A}} \mathbf{M}_{\mathrm{A} \mathrm{B}} (\mathbf{F}_{\mathrm{a}})_{\mathrm{B}}
\end{equation}
The term $\mathbf{f}_C(\mathbf{x},\mathbf{d})$ is then given by
\begin{align}
    \mathbf{f}_C(\mathbf{x},\mathbf{d}) & =  \underbrace{\begin{bmatrix}
    \frac{(\mathbf{t})_{\tau} (\mathbf{v}_k)_{\tau}}{||\mathbf{t}||_2 l_\Gamma} , \:
    \frac{(\mathbf{t}_{\perp})_{\tau} (\mathbf{v}_k)_{\tau}}{||\mathbf{t}_{\perp}||_2} , \:
    -(v_{\mathrm{k}})_{\tau, \mathrm{z}} , \:
    \frac{(F_{\mathrm{t}})_{\mathrm{\overline{\mathrm{A}}}, x}}{m_{\mathrm{a}}}, \:
    \frac{(F_{\mathrm{t}})_{\mathrm{\overline{\mathrm{A}}}, y}}{m_{\mathrm{a}} v_{\mathrm{a}} \cos{\gamma_{\mathrm{a}}}}, \:
    -\frac{(F_{\mathrm{t}})_{\mathrm{\overline{\mathrm{A}}}, z}}{m_{\mathrm{a}} v_{\mathrm{a}}}, \:
    0
\end{bmatrix}^T}_{=\mathbf{f}_{C1}(\mathbf{x},\mathbf{d}_{\mathrm{\mathrm{turb}}})} \\
& + \underbrace{\begin{bmatrix}
    0 , \:
    0 , \:
    0 , \:
    0, \:
    0, \:
    0, \:
d_{\Delta_{t}}
\end{bmatrix}^T}_{=\mathbf{f}_{C2}(\mathbf{x},d_{\Delta_{t}})} \\
& + \underbrace{\begin{bmatrix}
    0 , \:
    0 , \:
    0 , \:
\frac{(F_{\mathrm{g}, })_{\mathrm{\overline{\mathrm{A}}}, x}}{m_{\mathrm{a}}}, \:
\frac{(F_{\mathrm{g}})_{\mathrm{\overline{\mathrm{A}}}, y}}{m_{\mathrm{a}} v_{\mathrm{a}} \cos{\gamma_{\mathrm{a}}}}, \:
-\frac{(F_{\mathrm{g}})_{\mathrm{\overline{\mathrm{A}}}, z}}{m_{\mathrm{a}} v_{\mathrm{a}}}, \:
0
\end{bmatrix}^T}_{=\mathbf{f}_{C3}(\mathbf{x})}
\label{eqn:f_C}
\end{align}
where we adopt the same notation as before to denote the x, y, and z components of the forces in the $\overline{\mathrm{A}}$ frame and simplify the notation by omitting the dependence of $\mathbf{d}_{\mathrm{\mathrm{turb}}}$ and $\mathbf{x}$ in the tether and gravitational forces.

\section{HJ Reachability Analysis and Controller Synthesis} \label{sec:HJ}
\subsection{Problem Statement}
Having formulated the system dynamics, we are able to introduce the necessary concepts of HJ reachability analysis used for controller synthesis. The synthesized controller and subsequent hybrid control laws provide the second key contribution of this paper.

Let $\mathcal{R}$ be the reach set, the set of states that should be reached in a given time, and let $\mathcal{A}$ be the set of avoid states, the set of states that lead to a critical system failure (i.e., tether rupture). Then we can define the backward reachable set (BRS) as the set of states from which it is possible to reach the set $\mathcal{R}$ at the end of a given time interval with duration $T$ while guaranteeing never to enter the set $\mathcal{A}$ until then. Mathematically, let $\mathbf{x} \in \mathbb{R}^7$ be the system state defined in \eqref{eqn:states} evolving according to the ordinary differential equation
\begin{equation}
    \dot{\mathbf{x}}(t) = \mathbf{f}(\mathbf{x}(t), \mathbf{u}(t), \mathbf{d}(t)), \quad t \in [-T, 0], \mathbf{u} \in \mathcal{U}, \mathbf{d} \in \mathcal{D}
    \label{eqn:dynamics}
\end{equation}
Note that we treat time as negative consistent with \cite{Vertovec2022, Bansal2018}; this only simplifies some of the notation and implies that we start our system at time $-T$. The dynamics, $\mathbf{f}$, are assumed to be bounded and Lipschitz continuous in $\mathbf{x}$ and uniformly continuous in $\mathbf{u}$ and $\mathbf{d}$. Then, given the Lebesgue-measurable functions $\mathbf{u}(\cdot) \in \mathcal{U}$ and $\mathbf{d}(\cdot) \in \mathcal{D}$, the control and disturbance inputs, respectively, there exists a unique trajectory, $\boldsymbol\zeta$, solving \eqref{eqn:dynamics}, i.e.,
\begin{align*}
	\frac{\partial}{\partial t} \boldsymbol\zeta(t; \mathbf{x}_0, \mathbf{u}(\cdot), \mathbf{d}(\cdot)) &= 
	\mathbf{f}(\boldsymbol\zeta(t; \mathbf{x}_0, \mathbf{u}(\cdot), \mathbf{d}(\cdot)), \mathbf{u}(t), \mathbf{d}(t)) \: \forall t \in [-T,0]\\
	\boldsymbol\zeta(-T; \mathbf{x}_0, \mathbf{u}(\cdot), \mathbf{d}(\cdot)) &=  \mathbf{x}_0
\end{align*}
To capture the worst-case disturbance, we model the underlying control problem as a differential game of two players. Following \cite{Margellos2011}, we restrict the first player to play a nonanticipative strategy \cite{Varaiya1967, Evans1984}, which is a function $\boldsymbol\xi: \mathcal{D} \rightarrow \mathcal{U}$, such that for all $t \in [-T, 0]$ and for all $\mathbf{d}, \hat{\mathbf{d}} \in \mathcal{D}$, if $\mathbf{d}(\tau) = \hat{\mathbf{d}}(\tau)$ for almost every $\tau \in [-T, t]$, then $\boldsymbol\xi[\mathbf{d}](\tau) = \boldsymbol\xi[\hat{\mathbf{d}}](\tau)$ for almost every $\tau \in [-T, t]$. Furthermore, we use $\Sigma$ to denote the class of nonanticipative strategies.

Finally, we can define the BRS as
\begin{equation}
    \mathrm{BRS}_{\mathcal{A}, \mathcal{R}}(-T) = \Big\{ \mathbf{x} \in \mathbb{R}^7 \big| \exists \boldsymbol\xi(\cdot) \in \Sigma, \forall \mathbf{d}(\cdot) \in \mathcal{D}, 
    \big( \boldsymbol\zeta(0; \mathbf{x}, \boldsymbol\xi(\cdot), \mathbf{d}(\cdot)) \in \mathcal{R} \big) 
    \: \& \: (\forall \tau \in [-T, 0], \boldsymbol\zeta(\tau; \mathbf{x}, \boldsymbol\xi(\cdot), \mathbf{d}(\cdot)) \notin A) \Big\}
\end{equation}
In words, $\mathrm{BRS}_{\mathcal{A}, \mathcal{R}}(-T)$ is the set of states from which trajectories can start at $-T$, and there exists a choice for a non-anticipative strategy $\boldsymbol\xi$, such that for any disturbance strategy $\mathbf{d}$, the system state can reach $\mathcal{R}$ at the end of the horizon while avoiding $\mathcal{A}$ until then.

By defining the avoid states as the set of states that imply a tether rupture, and the reach set as the set of states that follow the optimal guidance strategy, the BRS is able to provide information about an impending critical failure. Furthermore, by computing the BRS, we are able to simultaneously find the optimal trajectory and control policy that allows for optimal flight while avoiding critical states.

Recall that the maximum force acting on the tether is given by the final segment attached to the aircraft. Thus, by setting the maximum allowed tether force as $F_{\mathrm{rupture}}$, we can define the Lipschitz continuous function $h(\mathbf{x}) = ||\mathbf{F}_{t}||_2 - F_{\mathrm{rupture}}$.
Then we can define the avoid set related to the superzero level-set of $h(\cdot)$ as $\mathcal{A} = \{ \mathbf{x} \in \mathbb{R}^7 | h(\mathbf{x}) > 0\}$.

Rather than defining our target set $\mathcal{R}$ as being the set of states that lie on the optimal tracking curve, we define our target set as all states that are aligned with the commanded flight direction. To this end, we can define Lipschitz continuous function $l(\cdot)$ by taking the signed distance functions of the course and path errors in the NED frame $l(\mathbf{x}) = \max\{\big| \gamma_{\mathrm{cmd}, O} -  (\gamma)_O \big|, \big| \chi_{\mathrm{cmd}, O} - (\chi)_O \big|\}$, where $(\gamma)_O$ and $(\chi)_O$ are the 5th and 6th states of our system transformed to the NED frame. The reach set is then defined by the subzero level-set of $l(\cdot)$, i.e., $\mathcal{R} = \{ \mathbf{x} \in \mathbb{R}^7 | l(\mathbf{x}) \leq 0\}$.

Using the definition of $\mathcal{A}$ and $\mathcal{R}$, as in \cite{Margellos2011}, it can be shown that $\mathrm{BRS}_{\mathcal{A}, \mathcal{R}}(-T) = \{ \mathbf{x} \in \mathbb{R}^7 | V(\mathbf{x},T) \leq 0\}$, where
\begin{equation}
	V(\mathbf{x},t) = \inf_{\boldsymbol\xi(\cdot) \in \Sigma} \sup_{\mathbf{d}(\cdot) \in \mathcal{D}} 
	\max \Big\{ l(\boldsymbol\zeta(0; \mathbf{x}, \boldsymbol\xi(\cdot), \mathbf{d}(\cdot)),
		\max_{\tau \in [-t, 0]} h(\boldsymbol\zeta(\tau; \mathbf{x}, \boldsymbol\xi(\cdot), \mathbf{d}(\cdot)) \Big\}
		\label{eqn:valuefcn}
\end{equation}
Furthermore, as in \cite{Margellos2011}, the value function in \eqref{eqn:valuefcn} is the unique continues viscosity solution of the following quasi-variational inequality
\begin{equation}
	\max\Big\{h(\mathbf{x}) - V(\mathbf{x},t), 
	\frac{\partial V(\mathbf{x},t)}{\partial t} + \sup_{\mathbf{d} \in D} \inf_{\mathbf{u} \in U} \frac{\partial V(\mathbf{x},t)}{\partial \mathbf{x}} f(\mathbf{x},\mathbf{u},\mathbf{d}) \Big\} = 0
	\label{eqn:HJB}
\end{equation}
with terminal condition $V(\mathbf{x},0) = \max\{h(\mathbf{x}), l(\mathbf{x})\}$ and where $D$ and $U$ are the compact sets of possible disturbance and control inputs, respectively as defined in \eqref{eqn:ud}. For numerical reasons, it has become common practice to solve \eqref{eqn:HJB} using level-set methods in combination with upwinding schemes such as essential non-oscillatory schemes \cite{Osher2002}. Alternative methods for solving \eqref{eqn:HJB}, such as using Deep Learning \cite{Bansal2020a} have been proposed, however, they lack many of the performance and safety guarantees we require.

\begin{figure}
 	\centering
	\includegraphics[width=0.45\textwidth]{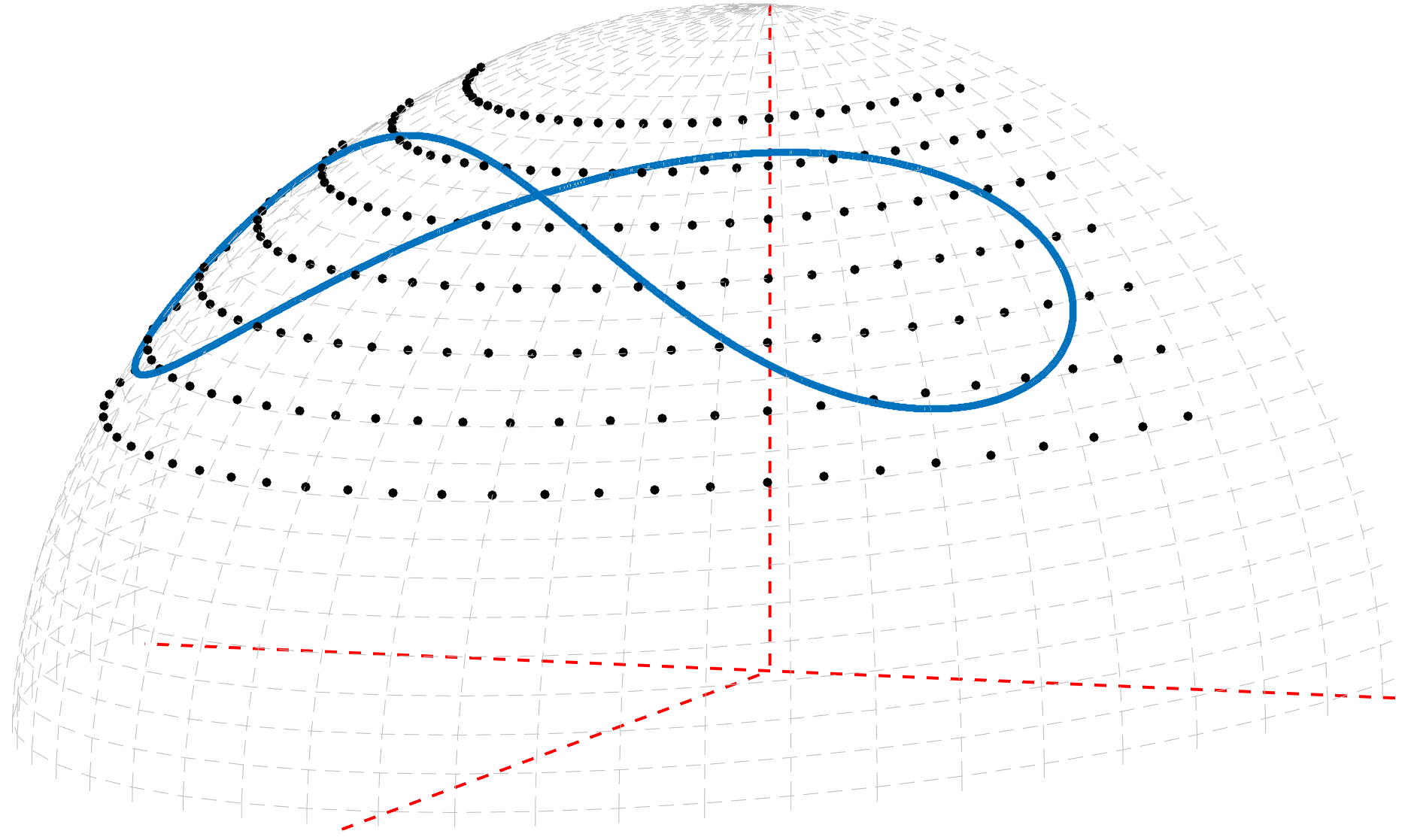}
	\includegraphics[width=0.45\textwidth]{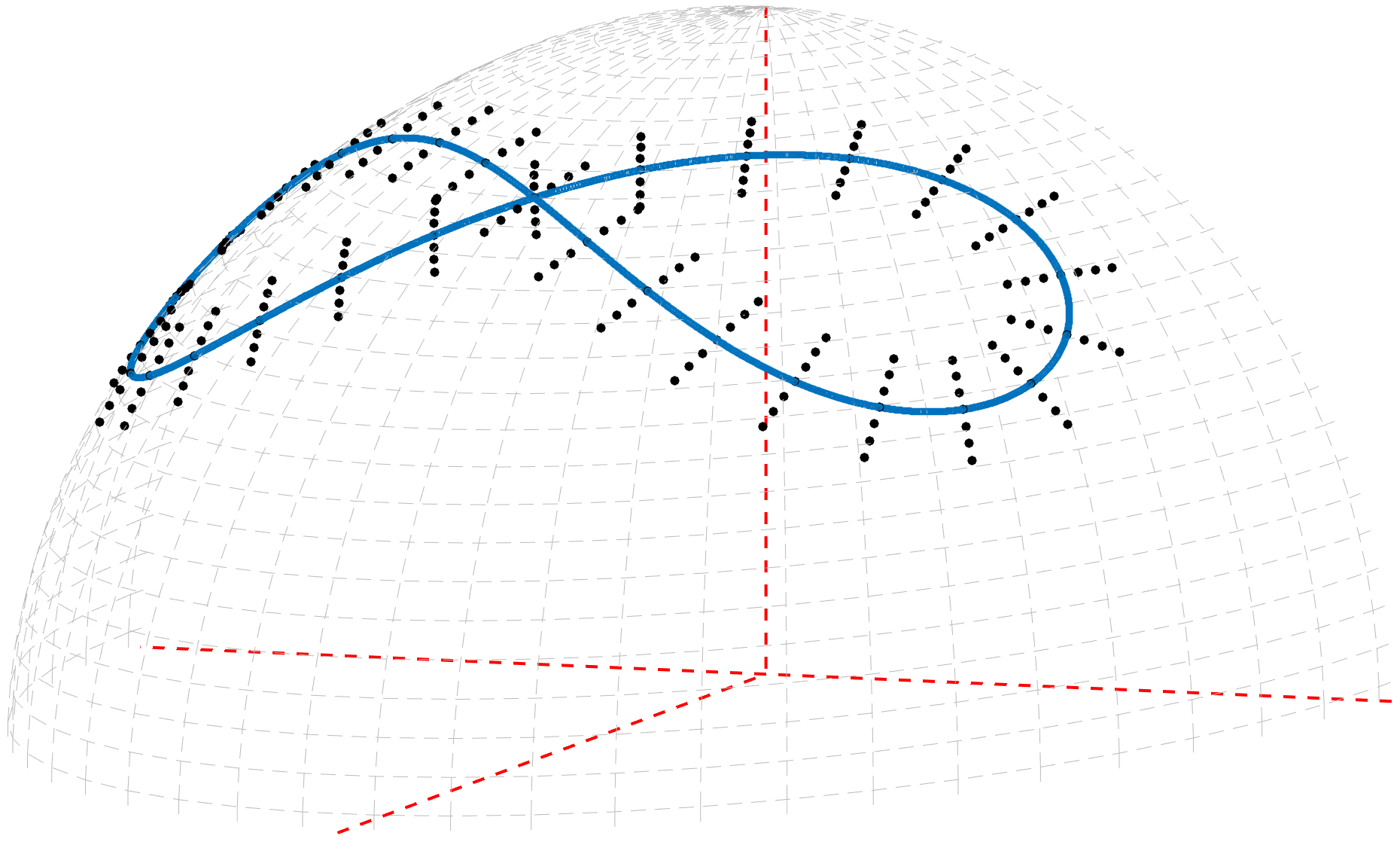}
	\caption{Comparison of a $31 \times 7$ grid in the NED frame vs the $\Gamma$ frame. The $\Gamma$ frame successfully captures only relevant nodes close to the optimal tracking curve.}
	\label{fig:gridding}
\end{figure}

To apply level-set methods to \eqref{eqn:HJB}, we are required to grid the state space, thus exponentially scaling the memory requirements with each additional state. This necessitates the need for the low dimensional model of the AWE system derived earlier in Section \ref{sec:modelforsafety}. Furthermore, for each grid node, we are required to evaluate the dynamics and calculate the derivative of the value function.
Therefore, to minimize computational overhead, we need to ensure we only consider grid nodes that are relevant during flight. To use conventional level-set methods, however, we are required to use an evenly spaced grid. As can be seen in Fig. \ref{fig:gridding}, gridding the position of the aircraft in the NED frame is inefficient compared to the new $\Gamma$ frame introduced in Section \ref{subsec:gammaframe}.  

\subsection{Optimal Control and Disturbance Inputs} \label{sec:optCTRL}
To solve \eqref{eqn:HJB}, we are required to find the optimal control and disturbance inputs that are the minimizers and maximizers, respectively, of $\sup_{\mathbf{d} \in D} \inf_{\mathbf{u} \in U} \frac{\partial V(\mathbf{x},t)}{\partial x} \mathbf{f}(\mathbf{x},\mathbf{u},\mathbf{d})$. 
To this end let us define the Hamiltonian of the system as 
\begin{equation}
    H(\mathbf{x},\mathbf{q}) = \max_{\mathbf{d} \in D} \min_{\mathbf{u} \in U} \mathbf{q}^T \mathbf{f}(\mathbf{x},\mathbf{u},\mathbf{d})
    \label{eqn:Ham0}
\end{equation}
where $\mathbf{q} = [q_1, \ldots, q_7] \in \mathbb{R}^7$ is the costate vector and since both $D$ and $U$ are compact, we are able to define the Hamiltonian using the $\max$ and $\min$ over $D$ and $U$, instead of the $\sup$ and $\inf$, respectively, i.e., the optimizers are achieved. To determine the optimizers in \eqref{eqn:Ham0} we utilize the separation of the dynamics presented in Section \ref{subsec:gammaframe}. This allows us to write the Hamiltonian as
\begin{equation}
    H(\mathbf{x},\mathbf{q}) = \min_{\mathbf{u} \in U} \mathbf{q}^T \hat{\mathbf{f}}(\mathbf{x},\mathbf{u}) + \max_{\mathbf{d} \in D} \mathbf{q}^T \mathbf{f}_C(\mathbf{x},\mathbf{d})
    \label{eqn:Ham1}
\end{equation}
separating the control from the disturbance inputs. We begin by finding the control inputs that minimize that Hamiltonian
\begin{equation}
    \mathbf{u}^* \in \argmin_{\mathbf{u} \in U} \mathbf{q}^T \hat{\mathbf{f}}(\mathbf{x},\mathbf{u}) 
     = \argmin_{\mathbf{u} \in U} \frac{1}{m_{\mathrm{a}}} \begin{bmatrix}
q_4 & \frac{q_5}{v_{\mathrm{a}} \cos{\gamma_{\mathrm{a}}}}  -\frac{q_6}{v_{\mathrm{a}}}
\end{bmatrix} \mathbf{M}_{\overline{\mathrm{A}} \mathrm{A}}(\mu) \mathbf{M}_{\mathrm{A} \mathrm{B}}(\alpha, \beta) (\mathbf{F}_{\mathrm{a}}(\alpha, v_a))_{\mathrm{B}}
\label{eqn:ustar}
\end{equation}

No analytic expression for \eqref{eqn:ustar} could be found, however, since we only need to know the optimal control inputs for a finite number of $v_a, \gamma, q_4, q_5$, and $q_6$ values, we grid the input space and evaluate \eqref{eqn:ustar} for discrete values of $\mu$ and $\alpha$ to find the near-optimal control inputs, i.e.,
\begin{equation}
    \mathbf{u}^* \in \argmin_{\substack{\alpha \in [\alpha_1, ..., \alpha_n], \mu \in [\mu_1, ..., \mu_m]}} v_{\mathrm{a}} q_4 \big(a(\alpha) \cos(\alpha) + b(\alpha) \sin(\alpha)\big)  +\big(\frac{q_5}{\cos{\gamma_{\mathrm{a}}}}\sin(\mu) + q_6 \cos(\mu))(b(\alpha) \cos(\alpha) - a(\alpha) \sin(\alpha)\big)
    \label{eqn:u_sol}
\end{equation}
where $a(\alpha)$ and $b(\alpha)$ are second order polynomials and $[\alpha_0, ..., \alpha_n]$ and $[\mu_1, ..., \mu_m]$ are discrete values of $\alpha$ and $\mu$, respectively. Notice that $u^*$ depends on the stat implicitly through the costate vector $q$.

Determining the control inputs that minimize the Hamiltonian by means of \eqref{eqn:u_sol}, we proceed to find the worst-case disturbances as the maximizer of the Hamiltonian, i.e., $
    \mathbf{d}^* \in \argmax_{\mathbf{d} \in D} \mathbf{q}^T \mathbf{f}_C(\mathbf{x},\mathbf{d}) 
     = \argmax_{d_{\Delta_{t}},\mathbf{d}_{\mathrm{\mathrm{turb}}}
    } \mathbf{q}^T \big(\mathbf{f}_{C1}(\mathbf{x},\mathbf{d}_{\mathrm{\mathrm{turb}}}) + \mathbf{f}_{C2}(\mathbf{x},d_{\Delta_{t}})\big)$
Due to this separable structure, we begin by computing the worst case disturbance, $d_{\Delta_{t}}$, which captures adverser winch control
\begin{equation}
    d_{\Delta_{t}}^* \in \argmax_{d_{\Delta_{t}} \in [-d_{\Delta_{t},\max}, d_{\Delta_{t},\max}]} \mathbf{q}^T \mathbf{f}_{C2}(\mathbf{x},d_{\Delta_{t}})
    \label{eqn:optdtether}
\end{equation}
where $d_{\Delta_{t},\max} = 0.005$ is considered the bound of $\dot{\Delta}_{t}$ as derived in Section \ref{subsec:simplifiedtether}. Solving \eqref{eqn:optdtether} yields $d_{\Delta_{t}}^* = -\mathrm{sign}(q_7) d_{\Delta_{t},\max}$.

For the computation of the disturbance vector $\mathbf{d}_{\mathrm{turb}}$, the worst case wind turbulence, we need to consider the effects the wind turbulence has on the position and heading of the aircraft. The wind turbulence naturally influences the wind velocity \eqref{eqn:windvelocity2} and subsequently the kinematic velocity of the aircraft \eqref{eqn:kinematicverlocity}. We thus begin by rewriting the kinematic velocity in the $\tau$ frame, whereby we replace the wind turbulence with the disturbance vector $\mathbf{d}_{\mathrm{turb}}$, i.e.,
\begin{equation}
    (\mathbf{v}_{\mathrm{k}})_{\tau} = \underbrace{\mathbf{M}_{\tau \mathrm{W}}\mathbf{M}_{\mathrm{W}\mathrm{O}} \Big(\mathbf{M}_{\mathrm{O}\overline{\mathrm{A}}} \begin{bmatrix}
     v_{\mathrm{a}} \\ 0 \\ 0
    \end{bmatrix}_{\mathrlap{\overline{\mathrm{A}}}} + (\mathbf{v}_{\mathrm{shear}})_{\mathrm{O}} \Big)}_{(\mathbf{v}_{\mathrm{k}, 0})_{\tau}} + \underbrace{\mathbf{M}_{\tau\mathrm{W}}(\mathbf{d}_{\mathrm{turb}})_{\mathrm{W}}}_{(\mathbf{d}_{\mathrm{turb}})_{\tau}}
    \label{eqn:kinematicdisturbance}
\end{equation}
This allows us to rewrite the position propagation in the NED frame as
\begin{align}
    \dot{\lambda} & = \underbrace{\frac{(v_{\mathrm{k}, 0})_{\tau, \mathrm{y}}}{\cos(\phi) h_{\tau}}}_{\dot{\lambda}_{0}} + \frac{(d_{\mathrm{turb}})_{\tau, \mathrm{y}}}{\cos(\phi) h_{\tau}} \label{eqn:positionpropagationwithturbulence1}
    \\
    \dot{\phi} & = \underbrace{\frac{(v_{\mathrm{k}, 0})_{\tau, \mathrm{x}}}{h_{\tau}}}_{\dot{\phi}_0} + \frac{(d_{\mathrm{turb}})_{\tau, \mathrm{x}}}{h_{\tau}}
    \label{eqn:positionpropagationwithturbulence2}
    \\
    \dot{h}_{\tau} & = \underbrace{-(v_{\mathrm{k}, 0})_{\tau, \mathrm{z}}}_{\dot{h}_{\tau,0}} -(d_{\mathrm{turb}})_{\tau, \mathrm{z}}
    \label{eqn:positionpropagationwithturbulence3}
\end{align}
where $(v_{\mathrm{k}, 0})_{\tau, \mathrm{x}}$, $(v_{\mathrm{k}, 0})_{\tau, \mathrm{y}}$ and $(v_{\mathrm{k}, 0})_{\tau, \mathrm{z}}$ denote the x, y, and z components of the base kinematic velocity in the $\tau$ frame and $(d_{\mathrm{turb}})_{\tau, \mathrm{x}}$, $(d_{\mathrm{turb}})_{\tau, \mathrm{y}}$ and $(d_{\mathrm{turb}})_{\tau, \mathrm{z}}$ denote the x, y and z components of the turbulence disturbance vector in the $\tau$ frame. 

\begin{figure}
 	\centering
	\includegraphics[width=0.7\textwidth]{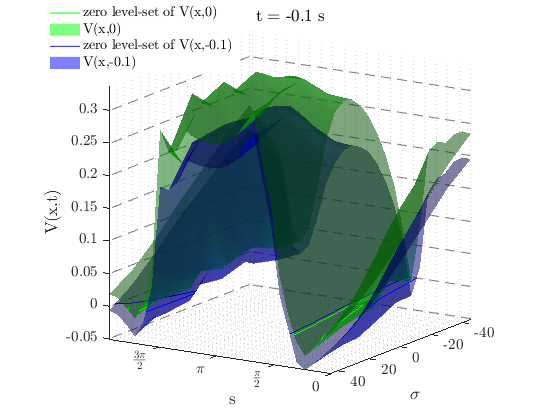}
	\caption{Visualization of the value function used for safety controller synthesis projected along 
	$[h_{\tau}=250m,v_a= 31m/s,\chi_a=-0.4470,\gamma_a = 0.5205,\Delta_{t}=0.0003m]$.}
	\label{fig:valufcn}
\end{figure}

Recall that the tether force \eqref{eqn:tetherforce} is also affected by the kinematic velocity of the aircraft through the difference between the velocity of the aircraft and the final tether segment, given by $\mathbf{s}_{v,n+1} = \mathbf{v}_{\mathrm{aircraft}} - \mathbf{v}_{n}$. To this end, let us rewrite $\mathbf{v}_{i}$ using \eqref{eqn:positionpropagationwithturbulence1}-\eqref{eqn:positionpropagationwithturbulence3}, to illustrate the dependence on $\mathbf{d}_{\mathrm{turb}}$,
\begin{multline}
    \mathbf{v}_{i} = \frac{i \dot{h}_{\tau,0}}{n+1} \begin{bmatrix}
    \cos(\phi) \cos(\lambda) \\
    \cos(\phi) \sin(\lambda) \\
    \sin(\phi)
    \end{bmatrix} - \frac{i (d_{\mathrm{turb}})_{\tau, \mathrm{z}}}{n+1} \begin{bmatrix}
    \cos(\phi) \cos(\lambda) \\
    \cos(\phi) \sin(\lambda) \\
    \sin(\phi)
    \end{bmatrix} - 
    \frac{i h_{\tau}}{n+1} \begin{bmatrix}
    \sin(\phi) \cos(\lambda) \dot{\phi}_0+ \cos(\phi) \sin(\lambda) \dot{\lambda}_0\\
    \sin(\phi) \sin(\lambda) \dot{\phi}_0 - \cos(\phi) \cos(\lambda) \dot{\lambda}_0\\
    -\cos(\phi) \dot{\phi}_0
    \end{bmatrix} - \\
    \frac{i (d_{\mathrm{turb}})_{\tau, \mathrm{x}}}{n+1} \begin{bmatrix}
    \sin(\phi) \cos(\lambda)\\
    \sin(\phi) \sin(\lambda)\\
    -\cos(\phi)
    \end{bmatrix} - 
    \frac{i (d_{\mathrm{turb}})_{\tau, \mathrm{y}}}{\cos(\phi) (n+1)} \begin{bmatrix}
    \cos(\phi) \sin(\lambda)\\
    -\cos(\phi) \cos(\lambda)\\
    0
    \end{bmatrix}
\end{multline}
Denoting the base difference (i.e., neglecting the the disturbance) as $\mathbf{s}_{v,n+1, 0}$, we can now rewrite $\mathbf{s}_{v,n+1}$ as 
\begin{equation}
    \mathbf{s}_{v,n+1} = \mathbf{s}_{v,n+1, 0} - \frac{ (d_{\mathrm{turb}})_{\tau, \mathrm{z}}}{n+1} \begin{bmatrix}
    \cos(\phi) \cos(\lambda) \\
    \cos(\phi) \sin(\lambda) \\
    \sin(\phi)
    \end{bmatrix} -
    \frac{(d_{\mathrm{turb}})_{\tau, \mathrm{x}}}{n+1} \begin{bmatrix}
    \sin(\phi) \cos(\lambda)\\
    \sin(\phi) \sin(\lambda)\\
    -\cos(\phi)
    \end{bmatrix} - 
    \frac{(d_{\mathrm{turb}})_{\tau, \mathrm{y}}}{\cos(\phi) (n+1)} \begin{bmatrix}
    \cos(\phi) \sin(\lambda)\\
    -\cos(\phi) \cos(\lambda)\\
    0
    \end{bmatrix}
\end{equation}
As can be seen, by choosing an $n$ sufficiently large, the effect of the disturbance on the tether force becomes negligible and we only need to consider the effects on the position propagation. Thus the worst-case turbulence disturbance can be calculated by 
\begin{equation}
    \mathbf{d}_{\mathrm{turb}}^* \in \argmax_{\substack{d_{\mathrm{turb}, \mathrm{x}} \in D_{\mathrm{turb}}, \\ d_{\mathrm{turb}, \mathrm{y}} \in D_{\mathrm{turb}} \\ d_{\mathrm{turb}, \mathrm{z}} \in D_{\mathrm{turb}}}} \begin{bmatrix}q_1 ,  q_2 ,  q_3\end{bmatrix} \begin{bmatrix}
    \dot{\lambda} ,  
    \dot{\phi} ,  
    \dot{h}_{\tau}
    \end{bmatrix}^T
\label{eqn:argminturb}
\end{equation}
where $D_{\mathrm{turb}} = [-d_{\mathrm{turb}, \max}, d_{\mathrm{turb}, \max}]$, with $d_{\mathrm{turb}, \max} = 4 \frac{m}{s}$, the possible range of turbulence velocities. The results in \eqref{eqn:positionpropagationwithturbulence1}-\eqref{eqn:positionpropagationwithturbulence3} utilize the disturbance in the $\tau$ frame, while we require the final result to be in the wind frame. To this end, we transform $(\mathbf{d}_{\mathrm{turb}})_{\tau}$ back into the wind frame and drop the additive term $(\mathbf{v}_{\mathrm{k}, 0})_{\tau}$, resulting in the final worst case turbulence disturbances that maximize the Hamiltonian
\begin{align*}
    d_{\mathrm{turb}, \mathrm{x}}^* & = -d_{\mathrm{turb}, \max} \cdot \mathrm{sign}\Big[ \sin(\phi) \cos(\lambda)\big( q_1 \frac{(t_{})_{\tau, \mathrm{x}}}{||\mathbf{t}||_2 l_\Gamma} 
    +q_2 \frac{(t_{\perp })_{\tau, \mathrm{x}}}{||\mathbf{t}_{\perp}||_2} \big) + \sin(\lambda) \big( q_1 \frac{(t_{})_{\tau, \mathrm{y}}}{||\mathbf{t}||_2 l_\Gamma} +q_2 \frac{(t_{\perp})_{\tau, \mathrm{y}}}{||\mathbf{t}_{\perp}||_2} \big) \\
    & + \cos(\phi) \cos(\lambda) \big( q_1 \frac{(t_{})_{\tau, \mathrm{z}}}{||\mathbf{t}||_2 l_\Gamma} +q_2 \frac{(t_{\perp})_{\tau, \mathrm{z}}}{||\mathbf{t}_{\perp}||_2}- q_3 \big) \Big] \\
    d_{\mathrm{turb}, \mathrm{y}}^* & = d_{\mathrm{turb}, \max} \cdot \mathrm{sign} \Big[ -\sin(\phi) \sin(\lambda) \big( q_1 \frac{(t_{})_{\tau, \mathrm{x}}}{||\mathbf{t}||_2 l_\Gamma} 
    +q_2 \frac{(t_{\perp})_{\tau, \mathrm{x}}}{||\mathbf{t}_{\perp}||_2} \big) + \cos(\lambda)\big( q_1 \frac{(t_{})_{\tau, \mathrm{y}}}{||\mathbf{t}||_2 l_\Gamma} +q_2 \frac{(t_{\perp})_{\tau, \mathrm{y}}}{||\mathbf{t}_{\perp}||_2} \big) \\
    &- \cos(\phi) \sin(\lambda) \big( q_1 \frac{(t_{})_{\tau, \mathrm{z}}}{||\mathbf{t}||_2 l_\Gamma} +q_2 \frac{(t_{\perp})_{\tau, \mathrm{z}}}{||\mathbf{t}_{\perp}||_2} - q_3 \big) \Big] \\
    d_{\mathrm{turb}, \mathrm{z}}^* & = d_{\mathrm{turb}, \max} \cdot \mathrm{sign} \Big[\cos(\phi) \big( q_1 \frac{(t_{})_{\tau, \mathrm{x}}}{||\mathbf{t}||_2 l_\Gamma} +q_2 \frac{(t_{\perp})_{\tau, \mathrm{x}}}{||\mathbf{t}_{\perp}||_2} \big) 
    - \sin(\phi) \big( q_1 \frac{(t)_{\tau,\mathrm{z}}}{||\mathbf{t}||_2 l_\Gamma} +q_2 \frac{(t_{\perp})_{\tau, \mathrm{z}}}{||\mathbf{t}_{\perp}||_2} - q_3 \big) \Big]
\end{align*}
With both the worst-case disturbance and optimal control inputs computed, the Hamiltonian can be calculated. To solve \eqref{eqn:HJB} we employ the Level Set Method toolbox of \cite{Mitchell2008} and begin by initializing the value function at $t=0$ with $V(\mathbf{x},0) = \max\{h(\mathbf{x}), l(\mathbf{x})\}$. As we solve the quasi-variational inequality \eqref{eqn:HJB}, we compute the optimal control inputs that form our control policy $\mathbf{u}(\mathbf{x},t)$. The evolution of the value function over the time period $[-0.1s, 0s]$ is shown in Fig. \ref{fig:valufcn}. Any state in the zero-sublevel set belongs to the BRS, i.e., the set of points that can reach $\mathcal{R}$ while avoiding $\mathcal{A}$ within $t$ units of time.

\subsection{Hybrid Control Setup} \label{subsec:hybridcontrol}
\begin{figure}
    \centering
    \begin{tikzpicture} [node distance = 3cm, on grid, auto]
    \node (q0) [state, initial, initial text = {start}] {$u_{\mathrm{NDI}}$};
    \node (q1) [state,right = of q0] {$u_{\mathrm{safety}}$};
     
    \path [-stealth, thick]
        (q0) edge [bend left] node[above] {$S_1(\mathbf{F}_{t})$}  (q1)
        (q1) edge [bend left] node[below] {$S_2(\mathbf{F}_{t})$}  (q0);
    \end{tikzpicture}
    \caption{State automaton of the hybrid control setup.}
    \label{fig:statemachine}
\end{figure}
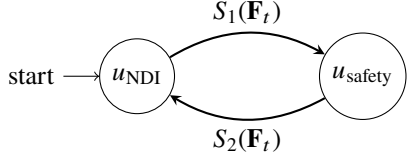

Let us refer to the safety and the NDI control laws as $u_{\mathrm{safety}}$ and $u_{\mathrm{NDI}}$ respectively. Then we can introduce the following switching laws
\begin{align}
    S_1(\mathbf{F}_{t}) & =
    (||\mathbf{F}_{t}||_2 \geq F_{\mathrm{rupture}}-30N) \quad \vee \quad 
     (||\mathbf{F}_{t}||_2 + \frac{\partial \overline{||\mathbf{F}_{t}||}_2}{\partial t} T \geq F_{\mathrm{rupture}}-50N) \label{eqn:switchinglaw1} \\
    S_2(\mathbf{F}_{t}) & =
    \neg \mathrm{S}_1(||\mathbf{F}_{t}||_2) \quad \wedge \quad 
    \Big[(||\mathbf{F}_{t}||_2 \leq F_{\mathrm{rupture}}-40N) \quad \vee \quad 
    (\frac{\partial \overline{||\mathbf{F}_{t}||}_2}{\partial t} \leq 0) \Big]
    \label{eqn:switchinglaw2}
\end{align}
where $T=0.1s$ is the time horizon used for the BRS computation and $\overline{||\mathbf{F}_{t}||}_2$ is the moving average of the tether force acting on the aircraft. The switching law needs to be tuned for a given AWE setup and influences the trade-off between safety and conservatism. Since the actuation delays are not accounted for in the safety control synthesis, belated switching to the safety controller will potentially not give the safety controller ample time to take the necessary action to avoid a critical system failure. The switching law chosen in this work is based on the current tether force as well as a prediction of the expected tether force, however, further extensions can be made to include the BRS as a maneuverability envelope as in \cite{Fisac2019, Herbert2021}. The condition in \ref{eqn:switchinglaw1} ensures that if the tether force comes within 30N of the critical force at which a rupture occurs, or, based on a linear extrapolation, the tether force will come within 50N of the critical tether force within the time horizon $T$, the safety controller will be activated. In turn, the condition \ref{eqn:switchinglaw2} ensures that we only switch back to the NDI controller if the tether force is decreasing or we are below 40N of the critical rupture force, thus preventing bang-bang control. Using the switching laws, the applied control action is determined using a simple state automaton as illustrated in Fig. \ref{fig:statemachine}.

\section{Simulation} \label{sec:simulation}
\subsection{Simulation Setup}
To simulate the full AWE setup, we extend the MATLAB Simulink framework presented in \cite{Rapp2021} and \cite{Eijkelhof2020}. Together with the subsequent discussion, the validation of the presented hybrid control setup constitutes our fourth key contribution of this paper. Some of the relevant simulation parameters are presented in Table \ref{tab:sim_param} and we have made the simulation code available \footnote{\url{https://github.com/nikovert/AWE_Simulation}}. 

\begin{table}[]
\centering
\caption{Simulation Parameters}
\label{tab:sim_param}
\begin{tabular}{lccc}
\hline \hline
Parameter     & Value  & Description                            \\ \hline
$F_{\mathrm{ref}}$     & 1600 N          & Force tracked by the winch controller  \\ 
$F_{\mathrm{rupture}}$ & 1870 N          & Force at which a tether rupture occurs  \\ 
$||W_{20}||$           & 9 $\frac{m}{s}$ & Measured wind sped at 20 feet            \\ 
$\xi$                  & $\pi$           & Wind direction                            \\
$n_{\mathrm{tether}}$    & 5               & Tether segments used  during simulation  \\
\hline \hline
\end{tabular}
\end{table}

In order to accurately capture the behavior of the tether and thus detect a tether rupture, we simulate the tether dynamics using $n = 5$, i.e., 30 states. The winch is also simulated as a separate subsystem with a PI controller regulating the reel-in and reel-out speed of the winch. Both the aircraft and the tether are affected by wind shear, which is modeled using the MATLAB Aerospace Toolbox \cite{AerospaceToolbox2021a}. In addition to the wind shear, the aircraft is also subject to wind turbulence, which is modeled using the continuous Dryden turbulence block \cite{AerospaceToolbox2021a}. 

The aircraft dynamics use the tether force computed by the full tether model as well as the wind velocity computed by the wind field model. To this end, we pass the altitude of the aircraft and the tether segments to the wind field model. Since the aircraft model simulates the dynamics using the $\tau$ and the $A$ frame, we need to convert the aircraft's longitude and latitude to the $\Gamma$ frame before computing the safety control input. We, therefore, add a Navigator module that does the necessary frame transformations. Furthermore, we use the Navigator module to compute optimal course and path angle rates $\dot{\chi}_{\mathrm{cmd}}$ and $\dot{\gamma}_{\mathrm{cmd}}$, respectively. The course and path angle rates are then used by the NDI controller to compute the pseudo control inputs $\nu_{\chi}$ and $\nu_{\gamma}$ based on the simplified aircraft dynamics. 

To acquire the safety control inputs, $u_{\mathrm{safety}}$, we solve \eqref{eqn:u_sol} using the numerically computed $\frac{\partial V(\mathbf{x}, T)}{\partial x}$ as our costate vector. Since calculating the derivatives of the value function is computationally intensive, it is impractical to compute $u_{\mathrm{safety}}$ during the simulation. To this end, we calculate the optimal safety control inputs offline using a high fidelity grid and save them on each grid point, using the saved values as a lookup table to be used during simulation.

Both control inputs, $u_{\mathrm{safety}}$ and $u_{\mathrm{NDI}}$ are passed to the control switch, which determines which control input to use based on the switching law derived in Section \ref{subsec:hybridcontrol}. The final control $u_{\mathrm{cmd}}$ is passed to the actuator, which simulates the actuator delays. The bank angle $\mu_{\mathrm{a}}$ and angle of attack $\alpha_{\mathrm{a}}$ are then passed to the aircraft model. To determine when to switch from traction to the retraction phase, we utilize the state machine, derived in \cite{Rapp2021}.

\subsection{Simulation Results}
\begin{figure}[ht]
 	\centering
	\includegraphics[scale=.23]{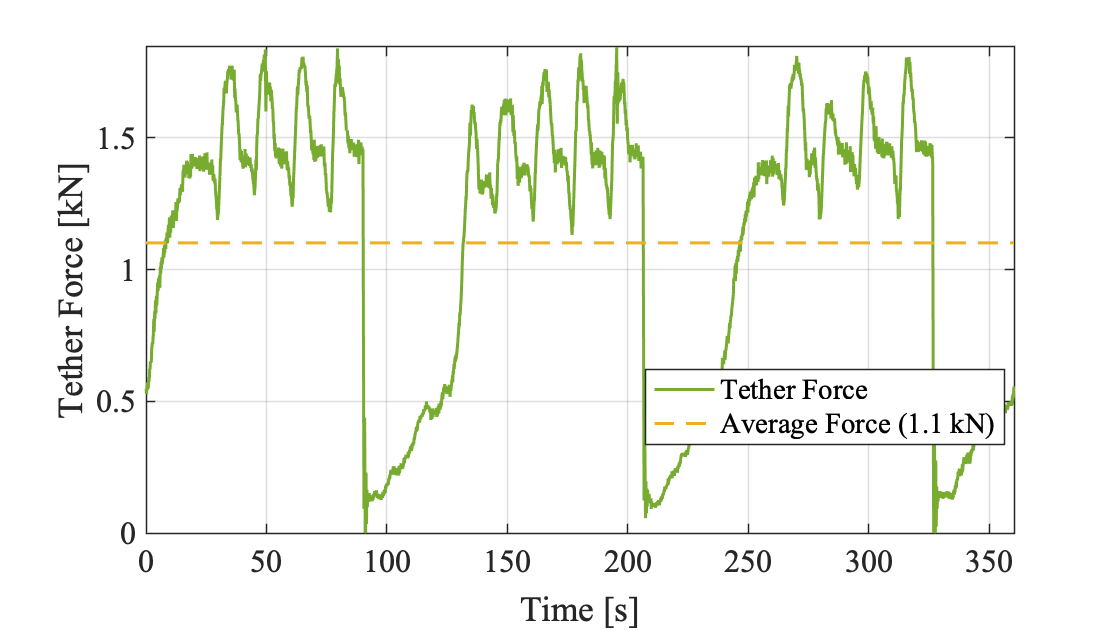}
	\caption{The tensile force of the tether acting on the aircraft during flight.}
	\label{fig:tether_force1}
\end{figure}

\begin{figure}[ht]
 	\centering
	\includegraphics[scale=.23]{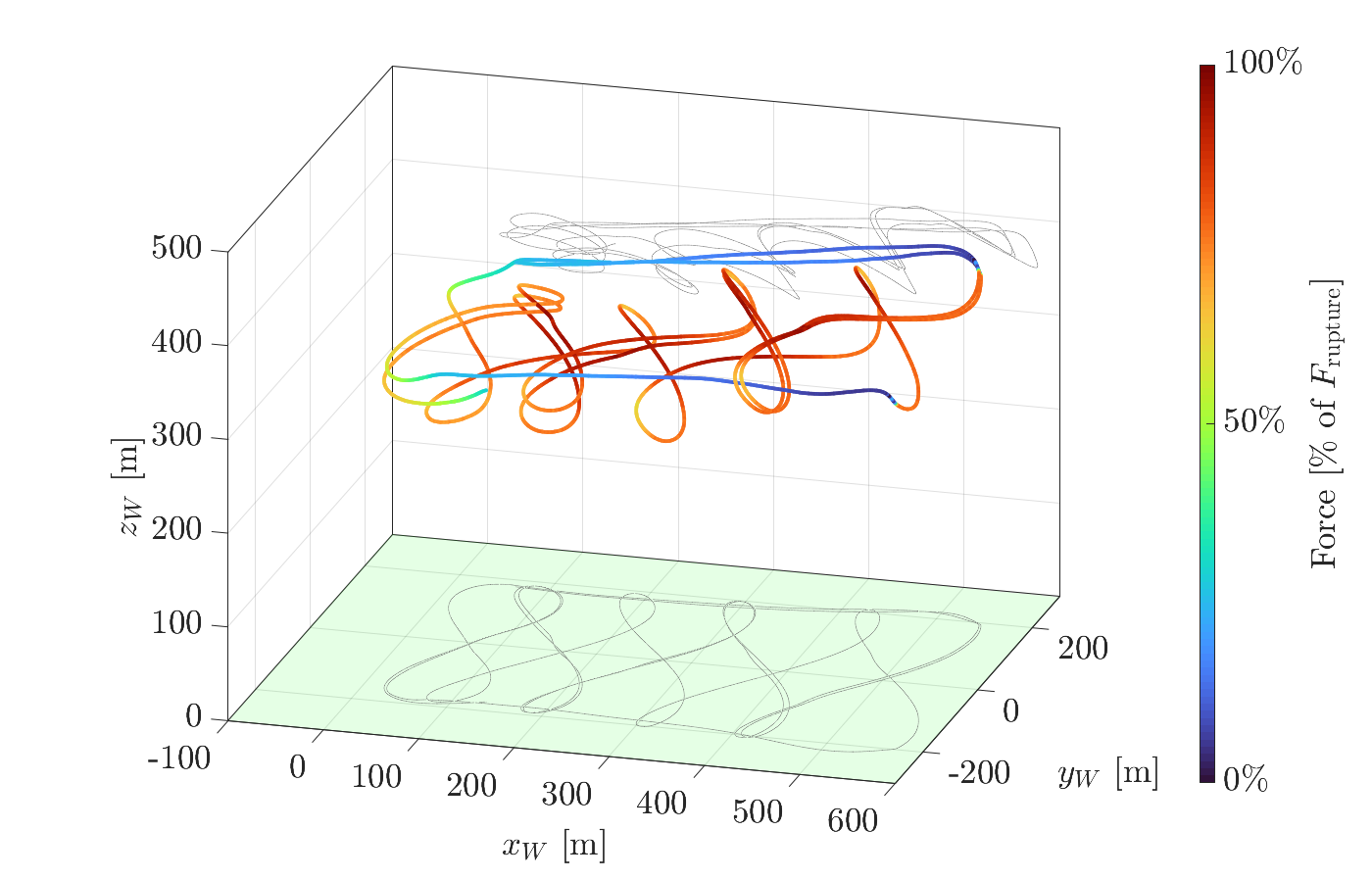}
	\caption{The tether force in $\%$ of the maximum allowed tether force before a rupture occurs.}
	\label{fig:tether_force2}
\end{figure}

\begin{figure}[ht]
 	\centering
	\includegraphics[scale=.23]{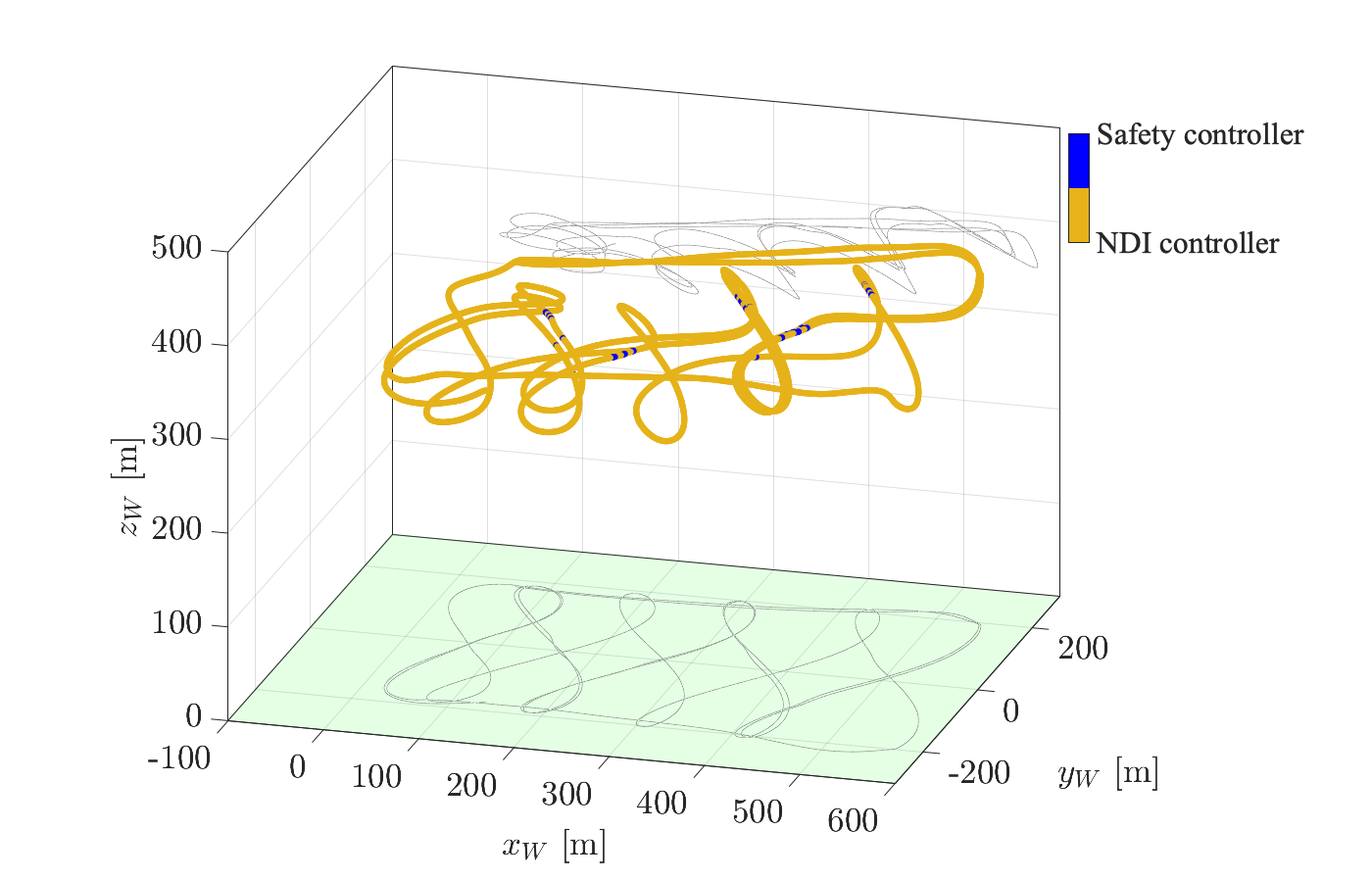}
	\caption{The safety controller only needs to be activated at particular times. For the remainder of the flight, a more power-optimal controller, such as the NDI controller can be employed.}
	\label{fig:tether_force3}
\end{figure}

To evaluate the performance of the safety controller, we run the simulation for multiple pumping cycles to determine if, at any point, the maximum allowed tether force is exceeded. The tether force of three pumping cycles as well as the switching behavior of the safety controller is shown in Fig. \ref{fig:tether_force1}, \ref{fig:tether_force2} and \ref{fig:tether_force3}. Notice how the tether force is able to remain high throughout the pumping cycle and only minimal interventions by the safety controller (blue segments in Fig. \ref{fig:tether_force3}) are necessary to prevent a tether rupture.

\begin{figure}[ht]
 	\centering
	\includegraphics[scale=.5]{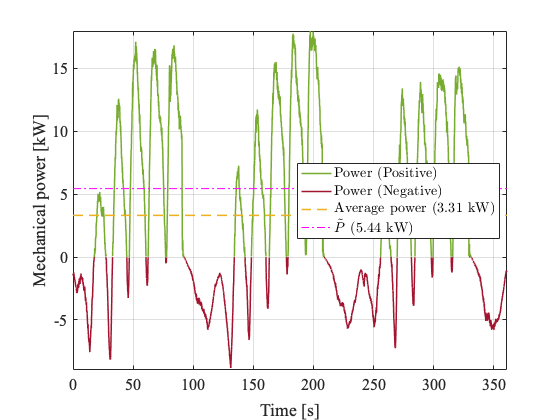}
	\caption{The power $P_{\mathrm{mech}}$ that is generated at the ground station during flight.}
	\label{fig:power1}
\end{figure}

\begin{figure}[ht]
 	\centering
	\includegraphics[scale=.23]{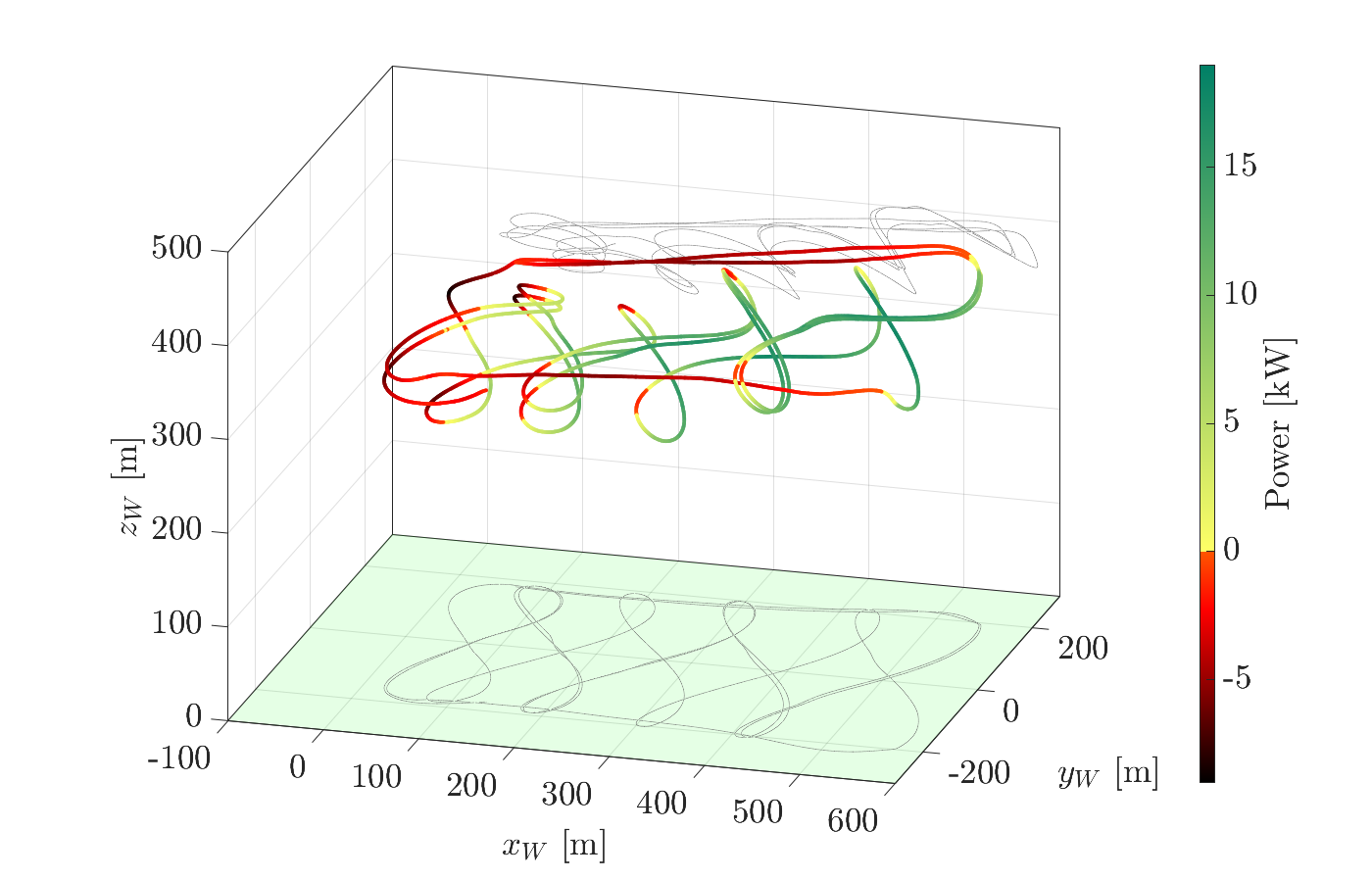}
	\caption{Generated power in relation to the aircraft's flight path. Negative (resp. positive) power indicates a reeling in (resp. out) by the winch thus power is exerted (resp. harvested).}
	\label{fig:power2}
\end{figure}

The power that is able to by harnessed by the ground station can be calculated by multiplying the tether force at the ground station by the reel-out speed, i.e., $P_{\mathrm{mech}} = \dot{\theta}_{\mathrm{W}} || \mathbf{F}_{\mathrm{w}} ||_2$.
In Fig. \ref{fig:power1} and \ref{fig:power2}, the power generation capabilities of the AWE system are shown. On average the chosen setup is able to produce 3.37 kW of power. 

To put the power generation into context, we derive a theoretical upper limit of the power that could be obtained through optimal flight based on the work of \cite{Costello2015}. We begin by calculating the maximum power harvesting factor, $\zeta$, based on the drag and lift coefficients presented in \cite{Rapp2019}. The power harvesting factor is a common metric used by both AWE applications as well as conventional wind turbines (not to be confused with the power coefficient commonly used for wind turbines)
\begin{equation}
    \zeta = \frac{4}{27} \frac{C_L^3}{(C_D+{C_D}_{\mathrm{tether}})^2}
\end{equation}
The power harvesting factor is unique for a given AWE setup and is used in \cite{Diehl2013} and \cite{Loyd1980} to calculate the theoretical limit $P_{\max}$. However, in order to calculate a tighter, more realistic upper bounds of the power that a given controller could achieve, we calculate an efficiency factor $e$ that can be multiplied by the theoretical upper bound of the power generation, $P_{\max}$. This theoretical upper bound is given by Theorem 1 in \cite{Costello2015}, $\tilde{P} = P_{\max} \underbrace{\cos^3(\gamma_0)}_{e}$, where $\gamma_0$ is the optimal angle between the aerodynamic force and the wind.

Based on the observation that the average aerodynamic force needs to balance the gravitational force acting on the aircraft as well as the force exerted on the aircraft by the tether (assuming that on average the system is not accelerating), we can state the following equality based on a 2-dimensional simplification
\begin{equation}
    ||\mathbf{F}_{a}||_2 \begin{bmatrix}
    \cos{\gamma_0} \\\sin{\gamma_0}
    \end{bmatrix} - ||\mathbf{F}_{t}||_2 \begin{bmatrix}
    \cos{\psi} \\\sin{\psi}
    \end{bmatrix} + \begin{bmatrix}
    ||\mathbf{F}_{\mathrm{drag}}||_2 \\ ||\mathbf{F}_{g}||_2
    \end{bmatrix} = 0
\end{equation}
where $\psi$ is the angle between the tether and the ground and $||\mathbf{F}_{\mathrm{drag}}||_2$ is the negated $x$ component of the aerodynamic force, $(\mathbf{F}_{a})_A$. On average, $\psi$ is equal to the rotation of the tracking curve, i.e., $\psi = \psi_0$. 

As in \cite{Costello2015}, we multiply both sides by $[\sin{\psi}, -\cos{\psi}]^T$, which cancels out the effect of the tether, allowing us to solve for $\gamma_0$, which in turn leads to the theoretical efficiency factor

\begin{equation}
    e = \cos^3 \Big( \psi_0 + \sin^{-1} \big( \frac{||\mathbf{F}_{\mathrm{drag}}||_2}{||\mathbf{F}_{a}||_2} \sin{\psi_0} + \frac{||\mathbf{F}_{g}||_2}{||\mathbf{F}_{a}||_2} \cos{\psi_0} \big) \Big)
\end{equation}

Finally, we calculate the theoretical maximum power that the hybrid control setup could obtain as
\begin{equation}
    \tilde{P} = e A_{\mathrm{eff}} \zeta \frac{1}{2} \rho ||\mathbf{v}_{\mathrm{W}}||_2^3
\end{equation}

When evaluated and averaged, this equates to a theoretical average power of 5.56 kW. Thus the hybrid control setup comes close to the upper bound that could be achieved.


\section{Conclusion}
\label{sec:conclusion}
We introduce a novel safety controller by deriving a low-dimensional simplification of the dynamics of an AWE system and applying HJ reachability analysis. By introducing switching conditions, the safety controller can be used in conjuncture with arbitrary controllers for safe and power-optimal flight. The hybrid control setup is simulated using a turbulence and wind shear model as well as actuation delays to simulate real-world environmental conditions. Using the hybrid control setup, we were able to successfully avoid a critical tether rupture while maintaining an average power output at a desired level. Future work will focus on improving the switching conditions between the safety controller and the NDI controller as well as incorporating actuator delays into the safety control framework.
\section*{Appendix}
\subsection*{Coordinate transformation matrices}
    For brevity, we only list the transformation matrices in one direction, as the inverse transformation is obtained by simply transposing the appropriate matrix, since all transformation matrices are orthogonal \cite{NairM2018}.
\begin{equation}
        \mathbf{M}_{\mathrm{A} \mathrm{B}} = 
    \begin{bmatrix}
    \cos{\alpha_{\mathrm{a}}} \cos{\beta_{\mathrm{a}}} & \sin{\beta_{\mathrm{a}}} & \sin{\alpha_{\mathrm{a}}} \cos{\beta_{\mathrm{a}}} \\
   -\cos{\alpha_{\mathrm{a}}} \sin{\beta_{\mathrm{a}}} & \cos{\beta_{\mathrm{a}}} & -\sin{\alpha_{\mathrm{a}}} \sin{\beta_{\mathrm{a}}} \\
   -\sin{\alpha_{\mathrm{a}}}                          & 0 & \cos{\alpha_{\mathrm{a}}}
    \end{bmatrix} , \quad
         \mathbf{M}_{\tau \mathrm{W}} = 
    \begin{bmatrix}
    -\sin{\phi} \cos{\lambda} & -\sin{\phi}\sin{\lambda} & \cos{\phi} \\
    -\sin{\lambda}            & \cos{\lambda} & 0 \\
    -\cos{\phi} \cos{\lambda} & -\cos{\phi}\sin{\lambda} & -\sin{\phi}
    \end{bmatrix}
\end{equation}
\begin{equation}
         \mathbf{M}_{\mathrm{O} \mathrm{W}} = 
    \begin{bmatrix}
     \cos{\xi} & \sin{\xi} & 0 \\
     \sin{\xi} & -\cos{\xi} & 0 \\
                0  &  0 & -1
    \end{bmatrix} , \quad
         \mathbf{M}_{\overline{\mathrm{A}} \mathrm{A}} = 
    \begin{bmatrix}
     1 & 0 & 0 \\
     0 & \cos{\mu_{\mathrm{a}}} & -\sin{\mu_{\mathrm{a}}} \\
     0 & \sin{\mu_{\mathrm{a}}} & \cos{\mu_{\mathrm{a}}}
    \end{bmatrix} , \quad
         \mathbf{M}_{\overline{\mathrm{A}} \mathrm{O}} = 
    \begin{bmatrix}
    \cos{\chi_{\mathrm{a}}} \cos{\gamma_{\mathrm{a}}} & \sin{\chi_{\mathrm{a}}} \cos{\gamma_{\mathrm{a}}} & -\sin{\gamma_{\mathrm{a}}} \\
   -\sin{\chi_{\mathrm{a}}} & \cos{\chi_{\mathrm{a}}} & 0 \\
   \cos{\chi_{\mathrm{a}}} \sin{\gamma_{\mathrm{a}}}   & \sin{\chi_{\mathrm{a}}} \sin{\gamma_{\mathrm{a}}}                     & \cos{\gamma_{\mathrm{a}}}
    \end{bmatrix}
\end{equation}

\section*{Acknowledgments}
The authors would like to acknowledge Dr. Sebastian Rapp, Dylan Eijkelhof, and Prof. Roland Schmehl for making their simulation framework available.

\bibliography{bibliography}

\begin{thebibliography}{33}
\newcommand{\enquote}[1]{``#1''}
\providecommand{\natexlab}[1]{#1}
\providecommand{\url}[1]{\texttt{#1}}
\providecommand{\urlprefix}{URL }
\expandafter\ifx\csname urlstyle\endcsname\relax
  \providecommand{\doi}[1]{\discretionary{}{}{}https://doi.org/#1}\else
  \providecommand{\doi}[1]{\discretionary{}{}{}\urlstyle{rm}\url{https://doi.org/#1}}\fi

\bibitem[{IEA(2021)}]{IEA_2021}
IEA, \enquote{Wind Power,} , November 2021.
\newblock \urlprefix\url{https://www.iea.org/reports/wind-power}.

\bibitem[{Schmehl(2018)}]{Schmehl2018}
Schmehl, R., \emph{Airborne Wind Energy: Advances in Technology Development and
  Research}, Springer, Singapore, 2018.

\bibitem[{Commission et~al.(2018)Commission, for Research, and
  Innovation}]{AWE2018}
Commission, E., for Research, D.-G., and Innovation, \emph{Study on challenges
  in the commercialisation of airborne wind energy systems}, Publications
  Office, 2018.
\newblock \doi{10.2777/87591}.

\bibitem[{Fagiano et~al.(2014)Fagiano, Zgraggen, Morari, and
  Khammash}]{Fagiano2014}
Fagiano, L., Zgraggen, A.~U., Morari, M., and Khammash, M., \enquote{{Automatic
  crosswind flight of tethered wings for airborne wind energy: Modeling,
  control design, and experimental results},} \emph{IEEE Transactions on
  Control Systems Technology}, Vol.~22, No.~4, 2014, pp. 1433--1447.
\newblock \doi{10.1109/TCST.2013.2279592}.

\bibitem[{Cherubini et~al.(2015)Cherubini, Papini, Vertechy, and
  Fontana}]{Cherubini2015}
Cherubini, A., Papini, A., Vertechy, R., and Fontana, M., \enquote{{Airborne
  Wind Energy Systems: A review of the technologies},} \emph{Renewable and
  Sustainable Energy Reviews}, Vol.~51, 2015, pp. 1461--1476.
\newblock \doi{10.1016/j.rser.2015.07.053}.

\bibitem[{Vermillion et~al.(2021)Vermillion, Cobb, Fagiano, Leuthold, Diehl,
  Smith, Wood, Rapp, Schmehl, Olinger, and Demetriou}]{Vermillion2021}
Vermillion, C., Cobb, M., Fagiano, L., Leuthold, R., Diehl, M., Smith, R.~S.,
  Wood, T.~A., Rapp, S., Schmehl, R., Olinger, D., and Demetriou, M.,
  \enquote{Electricity in the air: Insights from two decades of advanced
  control research and experimental flight testing of airborne wind energy
  systems,} \emph{Annual Reviews in Control}, Vol.~52, 2021, pp. 330--357.
\newblock \doi{https://doi.org/10.1016/j.arcontrol.2021.03.002}.

\bibitem[{Eijkelhof et~al.(2020)Eijkelhof, Rapp, Fasel, Gaunaa, and
  Schmehl}]{Eijkelhof2020}
Eijkelhof, D., Rapp, S., Fasel, U., Gaunaa, M., and Schmehl, R.,
  \enquote{{Reference Design and Simulation Framework of a Multi-Megawatt
  Airborne Wind Energy System},} \emph{Journal of Physics: Conference Series},
  Vol. 1618, No.~3, 2020.
\newblock \doi{10.1088/1742-6596/1618/3/032020}.

\bibitem[{Rapp(2021)}]{Rapp2021}
Rapp, S., \enquote{{Robust Automatic Pumping Cycle Operation of Airborne Wind
  Energy Systems Rapp,},} Ph.D. thesis, Delft University of Technology, 2021.
\newblock
  \doi{https://doi.org/10.4233/uuid:ab2adf33-ef5d-413c-b403-2cfb4f9b6bae}.

\bibitem[{Bansal et~al.(2017)Bansal, Chen, Herbert, and Tomlin}]{Bansal2018}
Bansal, S., Chen, M., Herbert, S., and Tomlin, C.~J., \enquote{Hamilton-Jacobi
  reachability: A brief overview and recent advances,} \emph{2017 IEEE 56th
  Annual Conference on Decision and Control (CDC)}, 2017, pp. 2242--2253.
\newblock \doi{10.1109/CDC.2017.8263977}.

\bibitem[{Vertovec et~al.(2021)Vertovec, Ober-Blöbaum, and
  Margellos}]{Vertovec2021}
Vertovec, N., Ober-Blöbaum, S., and Margellos, K., \enquote{Multi-objective
  minimum time optimal control for low-thrust trajectory design,} \emph{2021
  European Control Conference (ECC)}, 2021, pp. 1975--1980.
\newblock \doi{10.23919/ECC54610.2021.9654919}.

\bibitem[{Chen et~al.(2018)Chen, Herbert, Vashishtha, Bansal, and
  Tomlin}]{Chen2018}
Chen, M., Herbert, S.~L., Vashishtha, M.~S., Bansal, S., and Tomlin, C.~J.,
  \enquote{{Decomposition of Reachable Sets and Tubes for a Class of Nonlinear
  Systems},} \emph{IEEE Transactions on Automatic Control}, Vol.~63, No.~11,
  2018.
\newblock \doi{10.1109/TAC.2018.2797194}.

\bibitem[{Malz et~al.(2019)Malz, Koenemann, Sieberling, and Gros}]{Malz2019}
Malz, E., Koenemann, J., Sieberling, S., and Gros, S., \enquote{A reference
  model for airborne wind energy systems for optimization and control,}
  \emph{Renewable Energy}, Vol. 140, 2019, pp. 1004--1011.
\newblock \doi{https://doi.org/10.1016/j.renene.2019.03.111}.

\bibitem[{Rapp and Schmehl(2018)}]{Rapp2018}
Rapp, S., and Schmehl, R., \enquote{{Vertical takeoff and landing of flexible
  wing kite power systems},} \emph{Journal of Guidance, Control, and Dynamics},
  Vol.~41, No.~11, 2018, pp. 2386--2400.
\newblock \doi{10.2514/1.G003535}.

\bibitem[{Fechner et~al.(2015)Fechner, van~der Vlugt, Schreuder, and
  Schmehl}]{Fechner2015}
Fechner, U., van~der Vlugt, R., Schreuder, E., and Schmehl, R.,
  \enquote{{Dynamic model of a pumping kite power system},} \emph{Renewable
  Energy}, Vol.~83, 2015, pp. 705--716.
\newblock \doi{10.1016/j.renene.2015.04.028}.

\bibitem[{Jehle and Schmehl(2014)}]{Jehle2014}
Jehle, C., and Schmehl, R., \enquote{{Applied tracking control for kite power
  systems},} \emph{Journal of Guidance, Control, and Dynamics}, Vol.~37, No.~4,
  2014, pp. 1211--1222.
\newblock \doi{10.2514/1.62380}.

\bibitem[{Aer(2021a)}]{AerospaceToolbox2021a}
\enquote{MATLAB Aerospace Toolbox,} , 2021a.
\newblock The MathWorks, Natick, MA, USA.

\bibitem[{Mil(2012)}]{MilitaryHandbook}
\enquote{U.S. Military Handbook MIL-HDBK-1797B,} , April 2012.

\bibitem[{Diehl(2013)}]{Diehl2013}
Diehl, M., \enquote{{Airborne wind energy: Basic concepts and physical
  foundations},} \emph{Airborne Wind Energy}, edited by U.~Ahrens, M.~Diehl,
  and R.~Schmehl, Green Energy and Technology, Springer Berlin Heidelberg,
  Berlin, Heidelberg, 2013, pp. 3--22.
\newblock \doi{10.1007/978-3-642-39965-7}.

\bibitem[{Rapp et~al.(2019)Rapp, Schmehl, Oland, and Haas}]{Rapp2019}
Rapp, S., Schmehl, R., Oland, E., and Haas, T., \enquote{{Cascaded pumping
  cycle control for rigid wing airborne wind energy systems},} \emph{Journal of
  Guidance, Control, and Dynamics}, Vol.~42, No.~11, 2019, pp. 2456--2473.
\newblock \doi{10.2514/1.G004246}.

\bibitem[{Smith(1998)}]{Smith1998}
Smith, P., \enquote{A simplified approach to nonlinear dynamic inversion based
  flight control,} \emph{23rd Atmospheric Flight Mechanics Conference}, 1998,
  pp. 762--767.
\newblock \doi{10.2514/6.1998-4461}.

\bibitem[{Oprea(2007)}]{OpreaJohn2007}
Oprea, J., \emph{Differential geometry and its applications},
  2\textsuperscript{nd} ed., Mathematical Association of America, Washington,
  D.C., 2007.

\bibitem[{Vertovec et~al.(2022)Vertovec, Ober-Blöbaum, and
  Margellos}]{Vertovec2022}
Vertovec, N., Ober-Blöbaum, S., and Margellos, K., \enquote{Verification of
  safety critical control policies using kernel methods,} \emph{2022 European
  Control Conference (ECC)}, 2022, pp. 1870--1875.
\newblock \doi{10.23919/ECC55457.2022.9838224}.

\bibitem[{Margellos and Lygeros(2011)}]{Margellos2011}
Margellos, K., and Lygeros, J., \enquote{{Hamilton-jacobi formulation for
  reach-avoid differential games},} \emph{IEEE Transactions on Automatic
  Control}, Vol.~56, No.~8, 2011, pp. 1849--1861.
\newblock \doi{10.1109/TAC.2011.2105730}.

\bibitem[{Varaiya(1967)}]{Varaiya1967}
Varaiya, P.~P., \enquote{{On the Existence of Solutions to a Differential
  Game},} \emph{SIAM Journal on Control}, Vol.~5, No.~1, 1967, pp. 153--162.
\newblock \doi{10.1137/0305009}.

\bibitem[{Evans and Souganidis(1984)}]{Evans1984}
Evans, L., and Souganidis, P.~E., \enquote{{Differential games and
  representation formulas for solutions of Hamilton-Jacobi-Isaacs equations},}
  \emph{Indiana Univ. Math. J}, Vol.~33, No.~5, 1984, pp. 773--797.

\bibitem[{Osher and Fedkiw(2003)}]{Osher2002}
Osher, S., and Fedkiw, R., \emph{Level Set Methods and Dynamic Implicit
  Surfaces}, Applied Mathematical Sciences, Springer New York, 2003.

\bibitem[{Bansal and Tomlin(2021)}]{Bansal2020a}
Bansal, S., and Tomlin, C.~J., \enquote{DeepReach: A Deep Learning Approach to
  High-Dimensional Reachability,} \emph{2021 IEEE International Conference on
  Robotics and Automation (ICRA)}, 2021, pp. 1817--1824.
\newblock \doi{10.1109/ICRA48506.2021.9561949}.

\bibitem[{Mitchell(2008)}]{Mitchell2008}
Mitchell, I.~M., \enquote{{The flexible, extensible and efficient toolbox of
  level set methods},} \emph{Journal of Scientific Computing}, Vol.~35, No.
  2-3, 2008, pp. 300--329.
\newblock \doi{10.1007/s10915-007-9174-4}.

\bibitem[{Fisac et~al.(2019)Fisac, Akametalu, Zeilinger, Kaynama, Gillula, and
  Tomlin}]{Fisac2019}
Fisac, J.~F., Akametalu, A.~K., Zeilinger, M.~N., Kaynama, S., Gillula, J., and
  Tomlin, C.~J., \enquote{{A General Safety Framework for Learning-Based
  Control in Uncertain Robotic Systems},} \emph{IEEE Transactions on Automatic
  Control}, Vol.~64, No.~7, 2019, pp. 2737--2752.
\newblock \doi{10.1109/TAC.2018.2876389}.

\bibitem[{Herbert et~al.(2021)Herbert, Choi, Sanjeev, Gibson, Sreenath, and
  Tomlin}]{Herbert2021}
Herbert, S., Choi, J.~J., Sanjeev, S., Gibson, M., Sreenath, K., and Tomlin,
  C.~J., \enquote{Scalable Learning of Safety Guarantees for Autonomous Systems
  using Hamilton-Jacobi Reachability,} \emph{2021 IEEE International Conference
  on Robotics and Automation (ICRA)}, 2021, pp. 5914--5920.
\newblock \doi{10.1109/ICRA48506.2021.9561561}.

\bibitem[{Costello et~al.(2015)Costello, Costello, Fran{\c{c}}ois, and
  Bonvin}]{Costello2015}
Costello, S., Costello, C., Fran{\c{c}}ois, G., and Bonvin, D.,
  \enquote{{Analysis of the maximum efficiency of kite-power systems},}
  \emph{Journal of Renewable and Sustainable Energy}, Vol.~7, No.~5, 2015, pp.
  1--16.
\newblock \doi{10.1063/1.4931111}.

\bibitem[{Loyd(1980)}]{Loyd1980}
Loyd, M.~L., \enquote{{Crosswind Kite Power.}} \emph{Journal of energy},
  Vol.~4, No.~3, 1980, pp. 106--111.
\newblock \doi{10.2514/3.48021}.

\bibitem[{Nair and Singh(2018)}]{NairM2018}
Nair, M.~T., and Singh, A., \emph{Linear algebra}, Springer, Singapore, 2018.

\end{thebibliography}

\end{document}